\documentclass[journal]{IEEEtran}
\ifCLASSINFOpdf
\else
\fi
\usepackage{bbm}
\usepackage{verbatim}
\usepackage{graphicx}
\usepackage{xcolor,cite}
\usepackage{url}
\usepackage[cmex10]{amsmath}
\usepackage{amssymb}
\usepackage{amsmath}
\usepackage{algorithm, algorithmicx, algpseudocode}
\usepackage[caption=false,font=footnotesize]{subfig}
\usepackage{color}
\usepackage{cite}
\usepackage{epstopdf}
\usepackage{calc}
\usepackage{array}
\usepackage{soul}

\newcommand{\vr}[0]{\mathbf{r}}

\newcommand{\tx}[0]{\text{t}}

\newcommand{\rx}[0]{\text{r}}
\newcommand{\DA}[0]{\text{DA}}
\newcommand{\SA}[0]{\text{SA}}
\newcommand{\FH}[0]{\text{FH}}

\newcommand{\hermitian}[0]{\text{H}}

\makeatletter
\def\ps@IEEEtitlepagestyle{%
  \def\@oddfoot{\mycopyrightnotice}%
  \def\@evenfoot{}%
}
\def\mycopyrightnotice{%
  {\footnotesize This paper is a preprint (IEEE ``accepted'' status). IEEE copyright notice (\textcopyright~2018 IEEE.)\hfill}
  \gdef\mycopyrightnotice{}
}

\begin{document}
%
\title{Performance, Power, and Area Design Trade-offs in Millimeter-Wave Transmitter Beamforming Architectures}

\author{Han~Yan,~\IEEEmembership{Student~Member,~IEEE},~Sridhar~Ramesh,~Timothy~Gallagher,~\IEEEmembership{Member,~IEEE}, Curtis~Ling,~\IEEEmembership{Senior~Member,~IEEE}, and~Danijela~Cabric,~\IEEEmembership{Senior~Member,~IEEE}%
\thanks{Han Yan and Danijela Cabric are with the Electrical Engineering Department, University of California, Los Angeles, Los Angeles, CA 90095 (e-mail: yhaddint@ucla.edu; danijela@ee.ucla.edu).}
\thanks{Sridhar Ramesh, Timothy Gallagher, and Curtis Ling are with Maxlinear, Inc, Carlsbad, CA 92008. (e-mail: sramesh@maxlinear.com; tgallagher@maxlinear.com; cling@maxlinear.com)}}



\maketitle

\begin{abstract}

Millimeter wave (mmW) communications is viewed as the key enabler of 5G cellular networks due to vast spectrum availability that could boost peak rate and capacity. Due to increased propagation loss in mmW band, transceivers with massive antenna array are required to meet link budget, but their power consumption and cost become limiting factors for commercial systems. Radio designs based on hybrid digital and analog array architectures and the usage of radio frequency (RF) signal processing via phase shifters have emerged as potential solutions to improve radio energy efficiency and deliver performances close to conventional digital antenna arrays. In this paper, we provide an overview of the state-of-the-art mmW massive antenna array designs and comparison among three array architectures, namely digital array, partially-connected hybrid array (sub-array), and fully-connected hybrid array. The comparison of performance, power, and area for these three architectures is performed for three representative 5G downlink use cases, which cover a range of pre-beamforming signal-to-noise-ratios (SNR) and multiplexing regimes. This is the first study to comprehensively model and quantitatively analyze all design aspects and criteria including: 1) optimal linear precoder, 2) impact of quantization error in digital-to-analog converter (DAC) and phase shifters, 3) RF signal distribution network, 4) power and area estimation based on state-of-the-art mmW circuits including baseband digital precoding, digital signal distribution network, high-speed DACs, oscillators and mixers, phase shifters, RF signal distribution network, and power amplifiers. Our simulation results show that the fully-digital array is the most power and area efficient compared against optimal design for each architecture. Our analysis shows digital array benefits greatly from multi-user multiplexing. The analysis also reveals that sub-array is limited by reduced beamforming gain due to array partitioning, and system bottleneck of the fully-connected hybrid architecture is the excessively complicated and power hungry RF signal distribution network.
\end{abstract}


%
\IEEEpeerreviewmaketitle

%
%
\section{Introduction}
\label{sec:Introduction}

\IEEEPARstart{M}illimeter-wave (mmW) communications is a promising technology for the future fifth-generation (5G) cellular network \cite{6736746,5783993}. In the US, the Federal Communications Commission (FCC) has voted to adopt a new Upper Microwave Flexible Use service in the licensed bands, namely 28GHz (27.5-28.35GHz band), 37GHz (37-38.6GHz band), 39GHz (38.6-40GHz) with a total 3.85GHz bandwidth \cite{FCC2016}. The abundant spectrum facilitates key performance indicators (KPI) of 5G, including 10Gbps peak rate, 1000 times higher traffic throughput than the current cellular system \cite{Andrew:what5Gbe}. As shown in theory and measurements, mmW signals suffer higher free-space transmission loss \cite{6824746}, and is vulnerable to blockage \cite{6732923}. As a consequence, radios require beamforming (BF) with large antenna arrays at both base station (BS) and user equipment (UE) to combat severe propagation loss \cite{Rappaport:mmWavewillwork}. This makes reliable communication range short and as a consequence, mmW BSs will be deployed in an ultra-dense manner with inter-site distance in the order of hundreds of meters \cite{7010535,6736747}. Due to these facts, performance, energy, and cost efficiency in the future mmW base station (BS) radios become more important than ever before.

Implementation and deployment of transceiver arrays in sub-6GHz have shown great success. In the 4G Long Term Evolution Advanced (LTE-A) system, BS supports up to 8 antennas \cite{3GPP_lte-a} and arrays with even larger size are being actively prototyped \cite{Shepard:2012:APM:2348543.2348553} and will be soon available in the LTE-A PRO (the pre-5G standard). Those systems exclusively have digital array architecture based on a dedicated radio-frequency transceiver chain, with data converter and up/down-conversion, per each antenna, and rely on digital baseband for array processing. Many implementation challenges arise in scaling up array size \cite{6375940} by an order of magnitude or more required for mmW bands. System designers are also concerned about the high cost and power consumption in digital array architecture with massive number of RF-chains and ultra-wide processing bandwidth \cite{6894453}.

%
%
Recently, an emerging concept of hybrid array has been proposed. A hybrid array uses two stage array processing. The analog beamforming implemented with variable phase shifters (PS) provides beamforming gain and the digital beamforming in the baseband provides flexibility for multiplexing multiple user streams \cite{6979963,7010533}. As a result, hybrid arrays support an RF transceiver count which is smaller than the array size. Such an architecture intends to reduce the power and cost penalty due to numerous tranceivers. Based on the connectivity between RF-chain and antenna, there are two major variations, fully-connected hybrid array and partially connected hybrid array. Although both architectures were used for radar application \cite{MIT_evolution_of_array} and were introduced for telecommunication application as early as a decade ago \cite{1519678}, they have recently gained much attention for mmW radios. Signal processing techniques, including channel estimation and beamforming, using hybrid architecture have been comprehensively studied \cite{7400949}. Proposals for using hybrid architectures in mmW 5G have been considered in standardization organizations \cite{7959169}.


A handful comparative analyses exists for different mmW array architectures, with an emphasis on the signal process algorithms \cite{7959169,8030501,7342886,mmMAGIC_array}. Authors in \cite{7355304} discussed circuits design challenges in implementing energy-efficient digital arrays. The relationship between spectral efficiency (SE) and energy efficiency in partially-connected hybrid architecture is studied in \cite{7010533,7809100,7445130}. Works \cite{7961162,8030501} provided comparison among array architectures and concluded that hybrid architecture can achieve higher energy efficiency than fully digital ones in the regime of point-to-point communication. Future 5G system, however, will certainly use multiuser multiplexing to provide higher network throughput. Moreover, existing works did not study trade-offs among array size, transmit power, and specifications of key circuit blocks in the three architectures. However, system designers need to understand these trade-offs and hardware implications to develop energy and cost efficient mmW systems \cite{7080890}.

%
%

This work aims to fulfill this gap. We intend to compare different array architectures in a comprehensive manner by considering trade-offs among capacity, energy and area efficiency. Specifically, we compare array architectures based on the criterion of achieving same capacity. All design trade-offs are carefully considered in reaching most efficient design in all architectures which meets the requirement of typical 5G use cases. Power consumption, including analog processing energy and digital computation energy, and IC area are then compared based on state-of-the-art circuits. We provide several design insights on scaling laws and the bottlenecks in each architecture which allow us to predict  a trend for future wireless demands and technology scaling.

The paper is organized as follows. In Section~II, we briefly introduce emerging mmW array architectures and typical 5G use cases. In Section~III, we discuss design trade-offs in all array architectures and the designs used for comparison. In Section~IV, we study implementation issues in antenna arrays and their impact on different architectures. In Section~V, we present the state-of-the-art specifications of mmW beamforming circuits blocks and system level power consumption and IC area of the three architectures. This leads us to the general conclusions in Section~VI.

%
%
\section{Comparative Framework}
\label{sec:array architecture}
In this work, we focus on the comparison of transmitter antenna array architectures in a 5G mmW BS. We first introduce three commonly considered array architectures and summarize recent silicon implementations. Then, we describe the metrics used for comparison of the three architectures.


\begin{figure}[htp]

\subfloat[Block diagram of digital array.]{%
  \includegraphics[clip,width=1\columnwidth]{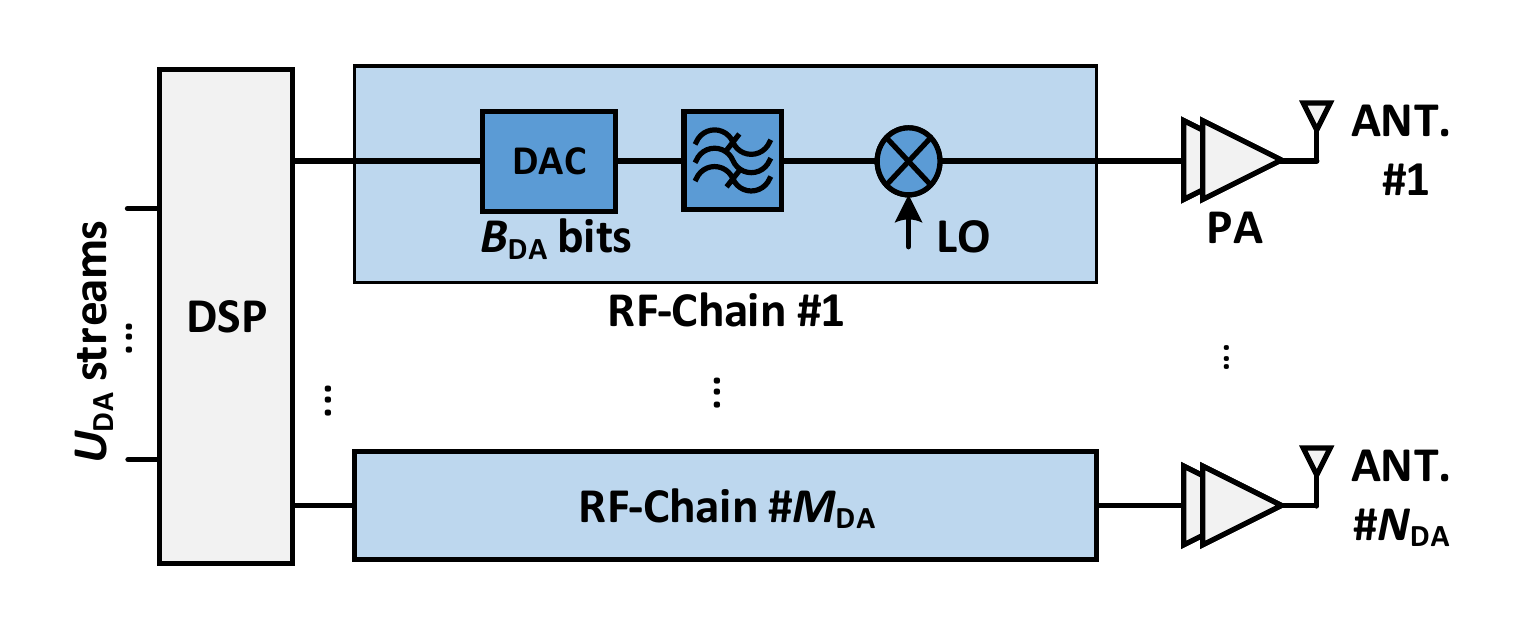}%
}

\subfloat[Block diagram of sub-array. Each RF-chain has the same structure as (a).]{%
  \includegraphics[clip,width=1\columnwidth]{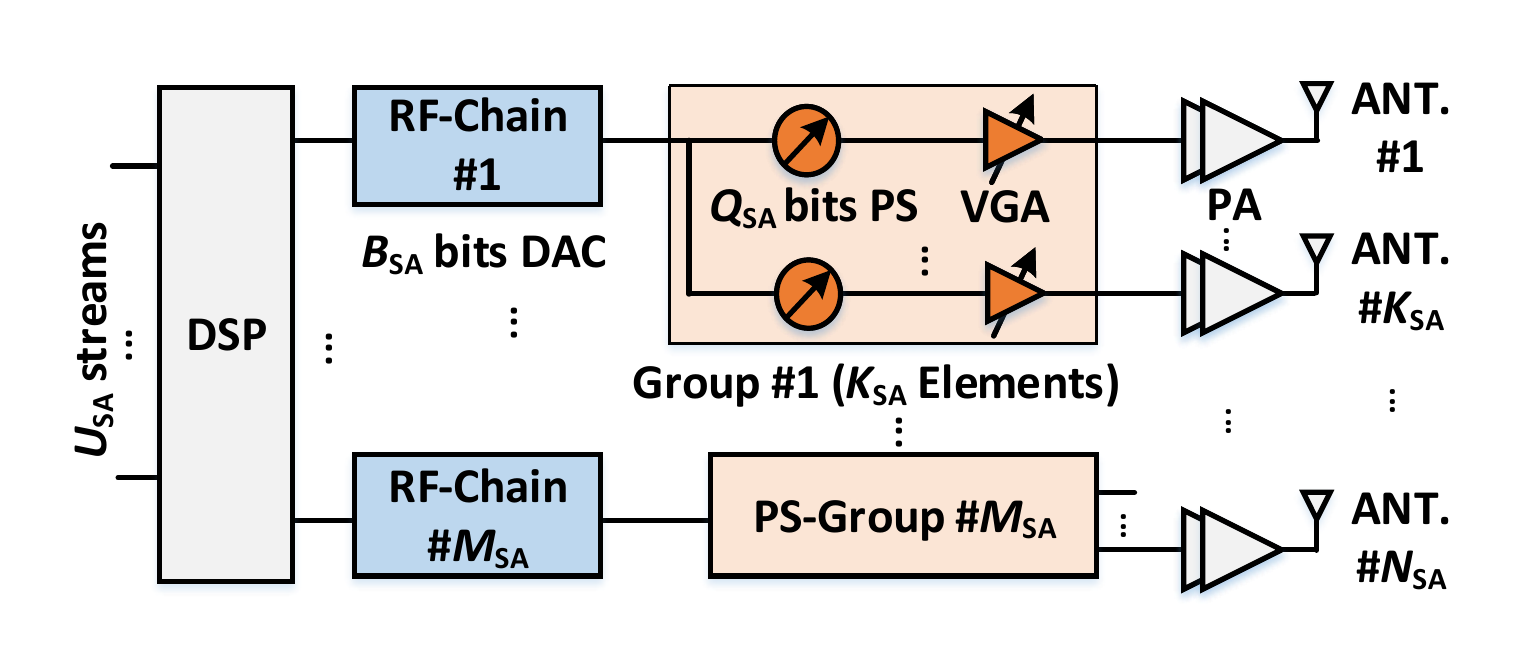}%
}

\subfloat[Block diagram of fully-connected hybrid array. Each RF-chain and PS group has the same structure as (b).]{%
  \includegraphics[clip,width=1\columnwidth]{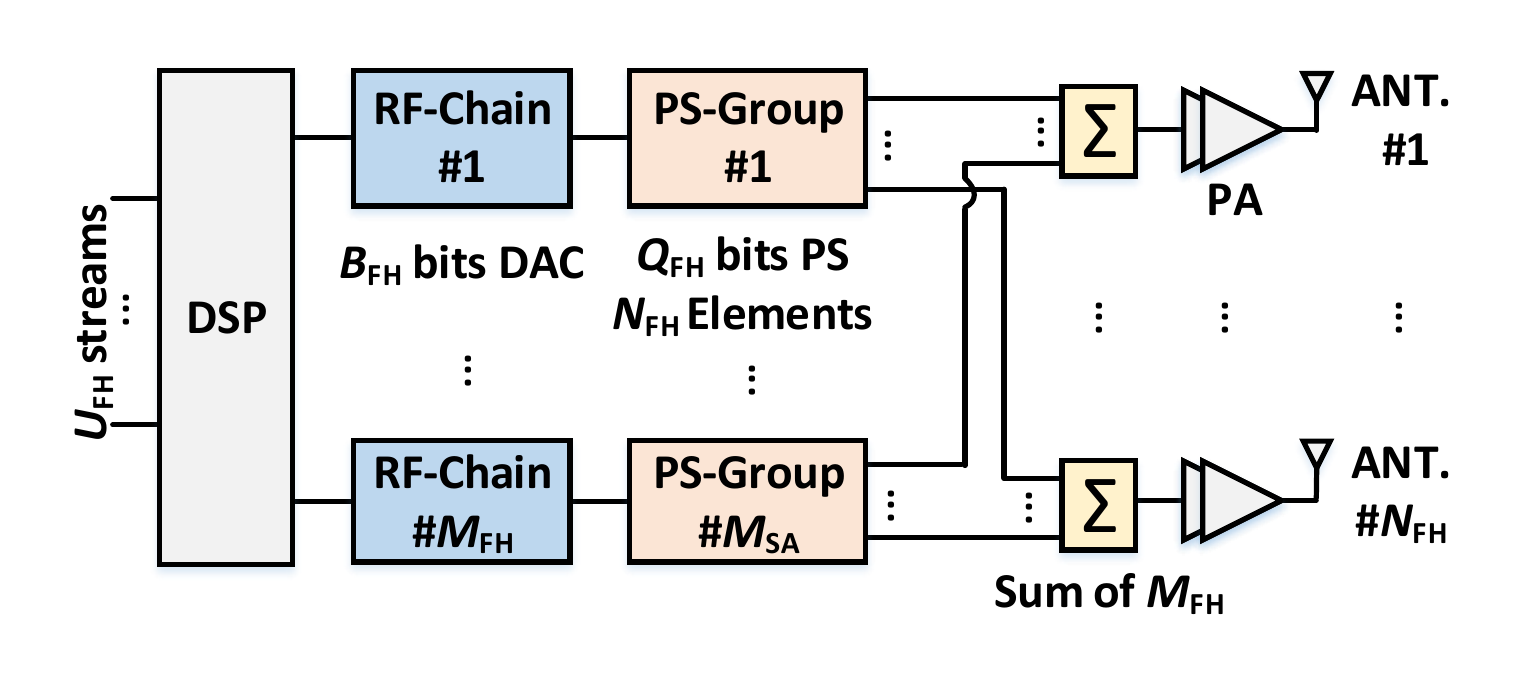}%
}

\caption{Three transmitter array architectures that are considered in this work.}
\label{fig:all_array_architecutre}
\end{figure}

%
%
\subsection{Array architectures}
\label{subsec:architecture_introduction}
There are three transmitter array architectures that are considered for adoption in 5G mmW system. Figure \ref{fig:all_array_architecutre} depicts block diagrams of digital array and two variations of hybrid array, partially-connected hybrid array (we denote it as sub-array in this work), and fully-connected hybrid array. Key design parameters for each architecture are:
\begin{itemize}
  \item Transmit power in all array elements: $P^{(\text{out})}$
  \item Number of antennas: $N$
  \item Number of RF-chains: $M$
  \item Number of simultaneous streams: $U$ $(U\leq M)$.
  \item Number of bits in digital-to-analog converter (DAC): $B$
  \item Number of bits in phase shifter: $Q$. This only applies to hybrid arrays.
\end{itemize}
In the rest of the paper, we use DA, SA and FH when referring to digital array architecture, sub-array and fully-connected hybrid array architecture, respectively. Mathematical symbols with subscript indicate parameters associated with the specific architecture, e.g., $N_{\DA}$ represents number of antennas in digital array. The main differences among three array architecture are:

\begin{itemize}
\item \textit{Digital Array:}
As shown in Figure~\ref{fig:all_array_architecutre}(a), $N_{\DA}$ antennas in DA are connected to $M_{\DA}$ RF-chains, i.e., $N_{\DA} = M_{\DA}$. The beamformer precoding occurs in the baseband (BB) digital signal processor (DSP).

\begin{table*}[t]
\caption{Silicon implementations of mmW array architectures}
\centering
\begin{tabular}{c|c|c|c|c|c|c|c|c|c}
\hline
\hline
Reference,& $\text{Archit-}$ & Freq. & Tx/Rx& Array& PA/ &LO&  Power Consumption per & Area per Array & Technology \tabularnewline
Year     &  $\text{ecture}$     & (GHz) &      & Size & LNA &   &  Array Element (mW)   &Element ($\text{mm}^2$) & \tabularnewline
\hline
\cite{7969030}, 2017 & FH & 25-30 & Rx & 8& - & $\checkmark$ & 30 (Rx) & 0.77 & 65nm CMOS \tabularnewline
\hline
\cite{7969017}, 2017 & SA & 28 & TRx & 2& - & - & 0 (both Tx and Rx) & 1.65 & 45nm CMOS\tabularnewline
\hline
\cite{7969012}, 2017 & SA & 28 & TRx & 4 & $\checkmark$ &-& 237.5 (Tx), 142.5 (Rx) & 1.23 & 65nm CMOS\tabularnewline
\hline
\cite{7797195}, 2017 & SA & 57-64 & Rx & 4& - &$\checkmark$ & 80 (Rx) & 0.65 & 65nm CMOS\tabularnewline
\hline
\cite{7417999}, 2016 & SA & 57-64 & TRx & 4& $\checkmark$ &$\checkmark$ & 167.5 (Tx), 107.8 (Rx) & 1.97 & 28nm CMOS\tabularnewline
\hline
\cite{6918546}, 2014 & SA & 57-64 & TRx & 16& $\checkmark$ &$\checkmark$ & 74.4 (Tx), 60 (Rx) & 2.07 & 40nm LP CMOS\tabularnewline
\hline
\cite{6464611}, 2013 & SA & 57-64 & TRx & 32& $\checkmark$ &$\checkmark$ & 37.5 (Tx), 26.6 (Rx) & 0.89 & 90nm CMOS\tabularnewline
\hline
\hline
\cite{7870294}, 2017 & SA & 28 & TRx & 32 & $\checkmark$ &$\checkmark$& 35.9 (Tx), 25.8 (Rx) & 5.18 & 0.13$\mu$m SiGe BiCMOS \tabularnewline
\hline
\cite{7337695}, 2015 & SA & 57-64 & Tx & 256& $\checkmark$ &- & 10.9 (Tx) & 6.79 & 0.18$\mu$m SiGe BiCMOS\tabularnewline
\hline
\cite{6867393}, 2014 & SA & 76-85 & TRx & 8 & $\checkmark$ &$\checkmark$& 118.74 (Tx), 143.8 (Rx) & 3.26 & 0.13$\mu$m SiGe BiCMOS\tabularnewline
\hline
\cite{6569608}, 2013& SA & 94 & TRx & 16& $\checkmark$ &$\checkmark$ & 181.25 (Tx), 156.25 (Rx) & 2.76 & 0.13$\mu$m SiGe BiCMOS \tabularnewline
\hline
\hline
\end{tabular}
\label{tab:array_prototypes}
\end{table*}

%
%
\item \textit{Sub-Array:}
SA consists of multiple phased arrays. As shown in Figure~\ref{fig:all_array_architecutre}(b), $N_{\SA}$ antennas are partitioned into $M_{\SA}$ group, each of which has one dedicated RF-chain, $K_{\SA}$ phase shifters (PS), variable gain amplifiers/attenuators (VGA), and power amplifiers (PAs). The array size, group number, and number of elements in a group follows relationship $N_{\SA} = M_{\SA}K_{\SA}$. Using phase shifters, each group can transmit a beam towards specific direction and SA is capable of transmitting/multiplexing up to $M_{\SA}$ simultaneous beams. When the required number of beams $U_{\SA}$ is smaller than $M_{\SA}$, multiple array groups can form a virtual group. The increased array size for that specific beam provides better beamforming performance, e.g., higher gain and narrower beam-width. DSP facilitates precoding multiple beams in the baseband. 



%
%
\item \textit{Fully-Connected Hybrid Array:} This architecture is also known as overlapped sub-array \cite{MIT_evolution_of_array}, multibeam active phased array \cite{1282136}, and high definition active antenna system \cite{blue_danube}. Similar to SA, the FH architecture uses phase shifters for analog beamforming and DSP for digital beamforming. However, FH has different connecting structures between RF-chains and phase shifters. As shown in Figure~\ref{fig:all_array_architecutre}(c), each of $M_{\FH}$ RF chains connects with all $N_{\FH}$ antennas via $N_{\text{FH}}$ phase shifters. Combiner networks are used to add $M_{\FH}$ RF signals before passing through the PAs. As a consequence, a total of $M_{\FH} N_{\FH}$ phase shifters are required in this architecture. FH is capable of transmitting up to $M_{\FH}$ simultaneous streams. 


\end{itemize}

Recent integrated circuits (IC) implementations of all three architectures are summarized in Table ~\ref{tab:array_prototypes}. Apart from array in 28GHz band, Table.~\ref{tab:array_prototypes} includes implementation in 60GHz band for mmW indoor access, mmW backhaul and \color{black} radar, because they share the same array architectures. Directly comparing array architectures from the table is difficult, because they use different silicon technology, and not all circuits components, e.g., local oscillator (LO) and associated up/down-conversion circuits, low noise amplifier (LNA), and PA, are integrated. It is worth noting that SA and FH architectures in Table.~\ref{tab:array_prototypes} implement phase shifters in the RF domain.  A comprehensive survey of phase shifter implementations is covered in \cite{poon2012}, including phase shifters in analog baseband, LO, and RF domain. Moreover, system level prototyping of 28GHz arrays together with field test can be found in \cite{7959169,MTT_64_TRx_array}. 

There are other architectures that have been recently proposed, e.g., switch based antenna array \cite{7370753} and lens antenna array \cite{6484896}. Due to the lack of implementation details available in the literature, we do not include quantitative analysis of them in this work.

%
%
\subsection{Comparison metrics under 5G use cases}

5G is characterized by a wide variety of use cases having different environments, communication distances, and performance requirements. Performance, in turn depends on connectivity density (defined as number of simultaneous connections for one wireless service operator in an given area), peak rate, and network traffic throughout. It is our vision that the mmW BS should be capable of using the same radio front-end arrays to handle various use cases and meet their demands. 


%
%

We choose three representative use cases \cite{mmMAGIC_requirement}: Dense Urban Mobile Broadband (MBB), 50+Mbps Everywhere, and Self-Backhauling. They cover different MIMO processing schemes of transmitter array.

%
%
\begin{itemize}
\item \textit{Dense Urban MBB:}
In dense urban area, large number of UEs require high-speed connections for applications like streaming, high-definition videos, and downloading files. According to 5G KPI requirement \cite{mmMAGIC_requirement}, the connection density is expected to be 150,000 connections per square kilometer, while the traffic throughput is up to 3.75Tbps/$\text{km}^2$ in such scenario. A typical 5G mmW BS deployment setting has inter-site distance (ISD) of 200m and each BS has 3 radio sectors \cite{3GPP_power}. With 850MHz spectrum at 28GHz band, the required SE in this use case is up to 58.8bps/Hz. Such a scenario often involves line-of-sight (LOS) environment and relatively good SNR is expected for each UEs so that SE greatly benefit from high multiplexing. We anticipate that at least 8 simultaneous streams are required\footnote{Till the time of writing, there is no specification for multiplexing in 5G mmW system. However, 8 streams are commonly used as assumption in the literature \cite{7878572,7914742}. Meanwhile, the next generation of 60GHz indoor wireless system also targets to use 8 spatial streams \cite{Rice_80211ay}.}.

\item \textit{50+Mbps Everywhere:}
mmW electromagnetic waves are extremely vulnerable to blockage. Despite this, BS in the 5G mmW network need to sustain baseline performance (up to 100Mbps data rate \cite{mmMAGIC_requirement}), even for those UEs under unfavorable propagation conditions. The 5G KPI requirement \cite{mmMAGIC_requirement} also indicated that the connection density is up to 2,500 connections per square kilometer. With the same BS deployment assumption as discussed in the previous use case, the required SE is 4.7bps/Hz. Due to a non-LOS (NLOS) environment, severe propagation loss exists and more than 20dB beamforming gain is required to close the link budget. Due to the requirement of high beamforming gain, we anticipated up to 8 simultaneous streams are adopted in this use case.

\item \textit{Self-Backhauling:}
To facilitate ultra-dense mmW BS deployment, BSs are required to connect to core network through a backhaul link. Since the large array allows interference isolation in the spatial domain, it is expected that 5G BS is capable of using the same spectrum for both access and backhauling, which is refereed as self-backhauling. Self-backhauling using radio for 5G access significantly reduces cost of setting up high-speed fiber. We consider a scenario where mmW BS transmits uplink data of its local network to a macro-BS receiver which connects to core network. With assumption of one macro-BS deployed in every square kilometer, the self-backhauling link has up to 707m communication distance \cite{Pi_Heath_backhaul_MCOM2016}. In this use case, LOS environment is assumed and 10Gbps rate is targeted by single data stream. 
\end{itemize}

For fair comparison of power consumption and area among array architectures, each array architecture has to deliver the same target SE. In Table~\ref{tab:link_budget_use_cases}, the system parameters and link budgets are summarized, with a set of possible data streams number $U$ and the corresponding signal to interference plus noise ratio (SINR) that reach SE objectives are also listed. In the Section~\ref{sec:design_parameters}, we study on the impact of design parameters on SE performance of different architectures and mainly focus on number of streams $U$, array size $N$ and required transmit power $P^{(\text{out})}$. The power consumption and hardware resources comparison are then presented based on state-of-the-art device specifications.

\begin{table}
\caption{Link Budget Estimation in Typical 5G Use Cases}
\centering
\begin{tabular}{|p{14mm}|p{3.5mm}|p{3.5mm}|p{3.5mm}|p{3.5mm}|p{3.5mm}|p{5mm}|p{12mm}|}
\hline
\textbf{Use Case} & \multicolumn{3}{c|}{\textbf{Dense-}} & \multicolumn{3}{c|}{\textbf{50+Mbps}} & \multicolumn{1}{c|}{\textbf{Self-}} \tabularnewline
  & \multicolumn{3}{c|}{\textbf{Urban MBB}} & \multicolumn{3}{c|}{\textbf{Everywhere}} & \multicolumn{1}{c|}{\textbf{Backhauling}} \tabularnewline
\hline
Channel& \multicolumn{3}{c|}{Umi-LOS} & \multicolumn{3}{c|}{Umi-NLOS}& \multicolumn{1}{c|}{Uma-LOS}\tabularnewline
\hline
Freq. $\text{[GHz]}$ & \multicolumn{3}{c|}{28}& \multicolumn{3}{c|}{28} & \multicolumn{1}{c|}{28}   \tabularnewline
\hline
BW $\text{[MHz]}$ & \multicolumn{3}{c|}{850} & \multicolumn{3}{c|}{850} & \multicolumn{1}{c|}{850}  \tabularnewline
\hline
Distance $\text{[m]}$ & \multicolumn{3}{c|}{100} & \multicolumn{3}{c|}{100} & \multicolumn{1}{c|}{707}\tabularnewline
\hline
$\text{Tx Power}$ $\text{[dBm]}$ & \multicolumn{3}{c|}{46.0} & \multicolumn{3}{c|}{46.0} & \multicolumn{1}{c|}{46.0} \tabularnewline
\hline
Tx Antenna Gain $\text{[dBi]}$ & \multicolumn{3}{c|}{3.0} & \multicolumn{3}{c|}{3.0} & \multicolumn{1}{c|}{3.0} \tabularnewline
\hline
$\text{Pathloss}^{\text{a}}$ $\text{[dB]}$ & \multicolumn{3}{c|}{104.4} & \multicolumn{3}{c|}{125.1} & \multicolumn{1}{c|}{118.3} \tabularnewline
\hline
$\text{Other Loss}^{\text{b}}$ $\text{[dB]}$ & \multicolumn{3}{c|}{12.7} & \multicolumn{3}{c|}{25.3} & \multicolumn{1}{c|}{17.0} \tabularnewline
\hline
$\text{Rx Gain}$ $\text{[dB]}$ & \multicolumn{3}{c|}{$12.0^{\text{c}}$} & \multicolumn{3}{c|}{$12.0^{\text{c}}$} & \multicolumn{1}{c|}{$27.1^{\text{d}}$} \tabularnewline
\hline
Rx NF $\text{[dB]}$ & \multicolumn{3}{c|}{10.0} & \multicolumn{3}{c|}{10.0} & \multicolumn{1}{c|}{10.0} \tabularnewline
\hline
Rx Noise $\text{[dBm]}$ & \multicolumn{3}{c|}{-74.7} & \multicolumn{3}{c|}{-74.7} & \multicolumn{1}{c|}{-74.7} \tabularnewline
\hline
SNR w/o Tx Array $\text{[dB]}$ &  \multicolumn{3}{c|}{18.7} &  \multicolumn{3}{c|}{-14.7} & \multicolumn{1}{c|}{15.5} \tabularnewline
\hline
Target SE $\text{[bps/Hz]}$& \multicolumn{3}{c|}{58.8}  & \multicolumn{3}{c|}{4.7} &  \multicolumn{1}{c|}{11.8}\tabularnewline
\hline
\hline
Simultaneous  Streams ($U$)& \multicolumn{1}{c|}{8} & \multicolumn{1}{c|}{16} & \multicolumn{1}{c|}{32} & \multicolumn{1}{c|}{2} & \multicolumn{1}{c|}{4} & \multicolumn{1}{c|}{8} &  \multicolumn{1}{c|}{1}\tabularnewline
\hline
Per-UE  $\text{SINR}^{\text{e}}$ $\text{[dB]}$& \multicolumn{1}{c|}{22.1} & \multicolumn{1}{c|}{10.7} &\multicolumn{1}{c|}{4.1} & \multicolumn{1}{c|}{6.2} & \multicolumn{1}{c|}{1.0} &\multicolumn{1}{c|}{-3.0}  & \multicolumn{1}{c|}{35.5}\tabularnewline
\hline
\multicolumn{8}{l}{a. Based on 3GPP model for above-6GHz band. \cite{3GPP_model}.}\tabularnewline
\multicolumn{8}{l}{b. Includes 3-sigma of shadowing loss and 25mm/h rain absorption \cite{6655399}.}\tabularnewline
\multicolumn{8}{l}{c. Based on 8 receiver antennas and 3dBi antenna gain.}\tabularnewline
\multicolumn{8}{l}{d. Based on 256 receiver antennas and 3dBi antenna gain.}\tabularnewline
\multicolumn{8}{l}{e. Based on equation $\text{SE} = U\log_2(1+\text{SINR})$.}\tabularnewline
\end{tabular}
\label{tab:link_budget_use_cases}
\end{table}
%
%
\section{Transmitter Array Design Parameters}
\label{sec:design_parameters}
In this section, we discuss the impact of array design parameters on the SE performance of multi-user multi-input multi-output (MU-MIMO) mmW system. We provide the design specification of components in array architectures to meet the SE requirement for each use case.

%
%
\subsection{System Model of mmW MU-MIMO}
We consider a mmW system where a BS of interest transmits data to multiple UEs in mmW access or a hub in mmW self-backhauling. Both transmitter and receiver are equipped with antenna array. Linear precoding techniques over flat fading channel are considered. In case of frequency selective channel, the precoding can be extended using orthogonal-frequency-division-multiplexing (OFDM) by considering per sub-carrier precoding. In the baseband equivalent model, the received symbol at the $u^{\text{th}}$ UE is denoted as 
\begin{align}
y_u = \mathbf{w}_u^{\hermitian} \mathbf{H}_u \mathbf{R} (\mathbf{B}  \mathbf{s}+\mathbf{z}_{\tx})+\mathbf{w}^{\hermitian}_{u} \mathbf{z}_{\rx}.
\label{eq:received_signal_model}
\end{align}
In the above equation, vector $\mathbf{s} = [s_1\cdots,s_U]$ contains the $U$ symbols. Matrix $\mathbf{H}_{u}$ is the MIMO channel between transmitter and $u^{\text{th}}$ UE receiver. Vector $\mathbf{w}_{u}$ represent the combining beamforming at the $u^{\text{th}}$ receiver. $\mathbf{B}$ and $\mathbf{R}$ denote the precoding scheme in the baseband and RF domain on the transmitter side, respectively. The transmit noise due to DAC quantization error is denoted as $\mathbf{z}_{\tx}$ and thermal noise at the receiver is $\mathbf{z}_{\rx}$. Operation $\mathbf{a}^{\hermitian}$ is the Hermitian transpose of $\mathbf{a}$.

In DA architecture, the precoding occurs entirely in digital baseband and therefore there is no analog processing, i.e., $\mathbf{R}_{\DA} = \mathbf{I}$. The digital precoder $\mathbf{B}_{\DA}$ has dimension $N_{\DA} \times U$.

In SA architecture, the digital precoder $\mathbf{B}_{\SA}$ has dimension $M_{\SA} \times U$ due to $M_{\SA}$ RF chains. The RF precoder $\mathbf{R}_{\SA}$ has dimension $N_\SA \times M_{\SA}$. Due to the fact that every $K_{\SA}$ of phase shifters connect to one RF-chain, $\mathbf{R}_{\SA}$ is a block diagonal matrix
\begin{align}
\mathbf{R}_{\text{SA}} = 
\text{diag}\left(\mathbf{r}_{\text{SA},1},\cdots,\mathbf{r}_{\text{SA},M}\right),
\end{align}
where column vector $\vr_{\text{SA},m}$ with length $K_{\SA}$ represents $K_{\SA}$ phase shifters that connect to the $m^{\text{th}}$ RF-chain. Each element of $\mathbf{r}_{\text{SA},m}$ has unit magnitude\footnote{In fact, analog precoding can be designed with both phase and magnitude tuning capability, which relaxes this constraint. The hardware aspect of phase shifter is discussed in Section V-C.}. We define the set $\mathcal{S}_m = \{(m-1)K_{\SA}+1,\cdots, mK_{\SA}\}$ that contains indices of array elements in the $m^{\text{th}}$ group. 

In FH architecture, the digital precoder $\mathbf{B}_{\FH}$ has dimension $M_{\FH} \times U$. The analog precoder matrix $\mathbf{R}_{\FH}$ has dimension $N_\FH \times M_{\FH}$ and its $m^{\text{th}}$ column $\mathbf{r}_{\text{FH},m}$ represents the phase shifting from $N_{\FH}$ phase shifters connected to the $m^{\text{th}}$ RF-chain, i.e.,
\begin{align}
\mathbf{R}_{\text{HF}} = 
\begin{bmatrix}%
\mathbf{r}_{\text{HF},1} & \mathbf{r}_{\text{HF},2} & \cdots & \mathbf{r}_{\text{HF},M_{\FH}} \\
\end{bmatrix}.
\end{align}
each element in $\mathbf{R}_{\FH}$ has unit magnitude.


We make the following assumptions. Firstly, the channel information $\mathbf{H}_u$ is known to both transmitter and receivers. A practical way of channel estimation can be found in \cite{7400949}. Secondly, each UE receiver is equipped with a phased array with only one RF-chain. As a consequence, BS assigns one data stream to each UE receiver. Thirdly, all receivers have the same pre-beamforming SNR and BS assigns equal power among data streams. Fourthly, the combining vector of each receiver $\mathbf{w}_u$ is chosen as the primary left eigenvector of channel matrix $\mathbf{H}_u$ after magnitude normalization in each element.

The SINR at the $u^{\text{th}}$ receiver array is denoted as
\begin{align}
\text{SINR}_u = \frac{\|g_u\|^2}{\sigma_{\text{n,rx}}^2+\sigma_{\text{n,tx}}^2+\sigma^2_{\text{int}}}
\label{eq:SINR_expression}
\end{align}
where the signal power gain $g_u$ is given by $g_u = \text{arg} \min_{g} \mathbb{E}\|y_u - g s_u\|^2 $. All signal, noise, and interference powers are relative powers, referenced to 46dBm transmit power based on Table~\ref{tab:link_budget_use_cases}. As a consequence, receiver thermal noise power $\mathbb{E}\|\mathbf{w}^{\hermitian}_u\mathbf{z}_{\rx}\|^2 = \sigma^2_{\text{n,rx}}$ is treated as constant in each use case. The multiuser interference is $\sigma^2_{\text{int}} = \mathbb{E}\|y_u - g_u s_u\|^2$.

In the remaining of the sections, we discuss how to design array parameters for each architecture to reach targeted SINR for three use cases.

%
%
\subsection{Array size and transmit power gain}
\label{subsec:array_size_design}
In principle, increased transmit power $P^{(\text{out})}$ and array size $N$ both improve signal power gain $g_u$ in (\ref{eq:SINR_expression}). Effectively, they provide higher equivalent isometric radiation power (EIRP) and help achieve target SINR from Table~\ref{tab:link_budget_use_cases}. 

In DA and FH, output power of each PA $P^{(\text{out})}/N$ is split into $U$ parts due to multiplexing and even power allocation. Thus each stream in each PA has output power $P^{(\text{out})}/(NU)$. The coherent summation of $N$-elements via beamforming provides $N^2$ times increased power. In SA, however, PAs are partitioned into groups to amplify different streams. For each stream, each PA element outputs $P_{\SA}^{(\text{out})}/N_{\SA}$, while the beamforming gain is $N^2_{\SA}/U^2$. As a consequence, maximum output signal power after beamforming in each architectures is
\begin{align}
G_{\DA} =  \frac{P_\DA^{(\text{out})}N_{\DA}}{U}, 
G_{\SA} =  \frac{P_\SA^{(\text{out})}N_{\SA}}{U^2}, 
G_{\FH} =  \frac{P_\FH^{(\text{out})}N_{\FH}}{U}.
\label{eq:maximum_gain}
\end{align}

It is clear that SA is in an disadvantage in terms of signal power gain. SA requires to use more array elements, output power, or both for the comparable output power to DA and FH architectures. 

%
%
\subsection{Precoder design}
\label{sec:precoder_design}
\begin{figure}[t]

\subfloat[In SA architecture, each beam is steered through a group of $N_{\SA}/U$ antenna elements.]{%
  \includegraphics[clip,width=1\columnwidth]{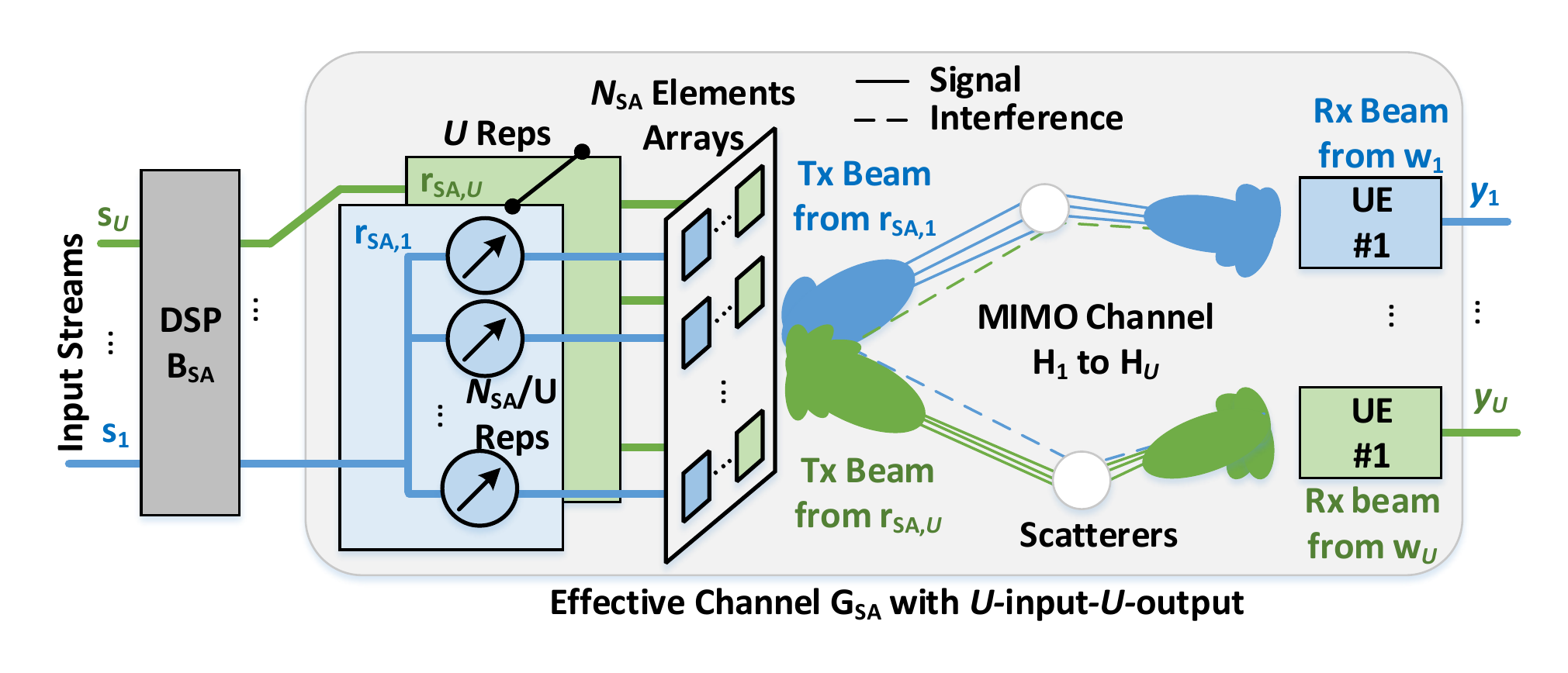}%
}

\subfloat[In FH architecture, each beam is steered through all $N_{\FH}$ antenna elements.]{%
  \includegraphics[clip,width=1\columnwidth]{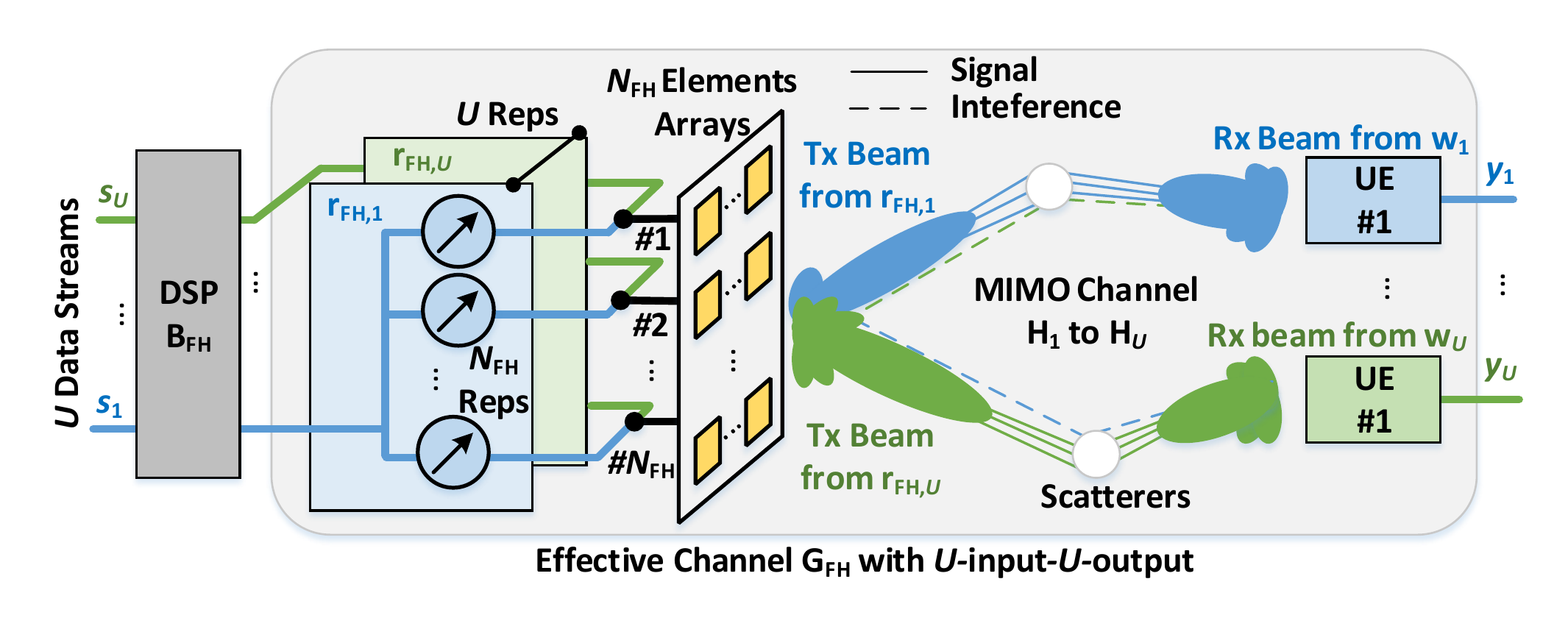}%
}

\caption{Two-stage precoding in SA and FH architectures. The analog precoder steers spatial beams towards intended receivers. The digital precoder uses regularized zero-forcing over effective channel to handle interference.}
\label{fig:illustration_precoding}
\end{figure}
Given maximum signal output power $G$, the the precoder determines the actual signal power $g_u$ and multiuser interference $\sigma^2_{\text{int}}$ in (\ref{eq:SINR_expression}). In this subsection, we discuss precoding techniques for three architectures.

In DA architecture, maximum ratio transmission (MRT) and zero-forcing (ZF) are two commonly used linear precoding approaches. The former maximizes the signal strength at destination and approaches maximum gain discussed in Section~\ref{subsec:architecture_introduction}, while the latter eliminates multiuser interference. It is commonly believed that because mmW signals suffer from severe propagation loss,  the interference is generally less troublesome than sub-6GHz systems. However, the interference from transmitted sidelobes, if not properly handled, can still affect the achievable rate at receivers. In this work, we propose to use regularized zero-forcing beamforming \cite{regularized_ZF}, where the introduced regularization coefficient $\alpha_{\DA}$ facilitates controlling both signal strength and interference at the receiver.
\begin{align}
\mathbf{B}_{\text{DA}} = \kappa_{\DA}\mathbf{G}_{\DA}^{\hermitian}(\mathbf{G}_{\DA}\mathbf{G}_{\DA}^{\hermitian}+\alpha_{\DA}\mathbf{I})^{-1},
\end{align}
In the above equation, $\mathbf{G}_{\DA}$ is the post-combining multiuser channel with the $u^{\text{th}}$ row as $\{\mathbf{G}_{\DA}\}_u = \mathbf{w}^{\hermitian}_u\mathbf{H}_u$. The regularization coefficient $\alpha_{\DA}$ controls the behavior of the precoder, i.e., MRT when it approaches positive infinity and ZF when it approaches zero. One can expect SINR maximization when $\alpha_{\DA}$ is selected to be the largest with constraint that $\sigma^2_{\text{int}} \ll \sigma^2_{\text{n,rx}}$. Power scaling parameter $\kappa_{\DA}$ is used to guarantee total transmit power constraint $\|\mathbf{B}_{\SA}\|^2=P_{\DA}^{(\text{out})}$.


Precoding approaches with SA and FH architectures are currently actively investigated by researchers and are mostly for systems where analog beamformer has phase-only tuning capability. The optimal hybrid precoding is a mixed integer programming problem and its optimal solution must be solved via potentially exhaustive search. Many sub-optimal methods have been proposed for near optimal performance, e.g., works in \cite{7160780} for FH architecture. In \cite{7160780}, the analog precoder is selected to point beams towards directions of intended receivers. The digital precoder is then used to handle associated interference among beams synthesized by phase shifters. In the following paragraphs regarding precoding algorithm for SA and FH, we adopt assumption of phase-only analog precoder.

In SA architecture, we propose to use the following approach as a modification of FH beamforming in \cite{7160780} and the scheme is illustrated in Figure~\ref{fig:illustration_precoding}(a). We first merge adjacent $M_{\SA}/U$ phase shifter groups in SA into one virtual group. It leads to $N_{\SA}/U$ array elements within each virtual group in an ideal scenario\footnote{Ideal scenario is defined when the ratio $M_{\SA}/U$ is an integer. Using a reduced number of arrays can be used when it is not valid, but this scenario is not considered for simplicity.}. The input signal of RF-chains within a virtual group are exactly the same. Let us denote set $\mathcal{V}_u$ as one that contains index of physical array groups within the $u^{\text{th}}$ virtual group. The analog beamformer is chosen to synthesize beams towards primary propagation direction to $U$ receivers
\begin{align}
\mathbf{r}_{\SA,m} = \text{exp}\left[j \angle \left(\{\mathbf{H}^{\hermitian}_u\mathbf{w}_u\}_{\mathcal{S}_m}\right) \right], m\in \mathcal{V}_u.
\end{align}
In the above equation, $\angle(\{\mathbf{a}\}_{\mathcal{S}_m})$ selects elements from vector $\mathbf{a}$ according to indices from set $\mathcal{S}_m$ and finds phases of selected elements. Let us denote the effective channel as $\mathbf{G}_{\SA}$ which contains the effect of receiver combiner and RF precoder in multiuser channel. The $m^{\text{th}}$ row is defined as $\{\mathbf{G}_{\text{SA}}\}_m = \mathbf{w}_m^{\hermitian} \mathbf{H}_m \mathbf{R}_{\SA}$. Note the effective channel $\mathbf{G}_{\SA}$ is the channel between digitally precoded stream and UEs. As a consequence, the digital precoding problem in SA can be solved in the regularized-ZF framework
\begin{align}
\mathbf{B}_{\SA} = \kappa_{\SA}\mathbf{G}_{\text{SA}}^{\hermitian}(\mathbf{G}_{\text{SA}}
\mathbf{G}_{\text{SA}}^{\hermitian}+\alpha_{\SA} \mathbf{I})^{-1}
\end{align}
The power scaling coefficient $\kappa_{\SA}$ is used to meet total output power constraint, i.e., $\|\mathbf{R}_{\SA}\mathbf{B}_{\SA}\|^2 = P_{\SA}^{(\text{out})}$. Similar to precoding in the digital array, the regularization coefficient $\alpha_{\SA}$ is chosen to maximize SINR.

The precoding scheme in FH architecture is illustrated in Figure~\ref{fig:illustration_precoding}(b). 
Only $U$ out of $M_{\FH}$ RF-chains are turned-on to provides $U$ streams. Without loss of generality, the first $U$ RF-chains are active and the analog precoder is
\begin{align}
\mathbf{r}_{\text{FH},u} = \text{exp}\left[j \angle (\mathbf{H}^{\hermitian}_u\mathbf{w}_u) \right], u\leq U.
\end{align}
The digital precoder in FH is a regularized zero-forcing over $\mathbf{G}_{\text{FH}}$, the effective channel that contains the receiver combining and RF precoding in the multiuser channel
\begin{align}
\mathbf{B}_{\FH} = \kappa_{\FH}\mathbf{G}_{\text{FH}}^{\hermitian}(\mathbf{G}_{\text{FH}}
\mathbf{G}_{\text{FH}}^{\hermitian}+\alpha_{\FH} \mathbf{I})^{-1},
\end{align}
The $u^{\text{th}}$ row is defined as $\{\mathbf{G}_{\text{FH}}\}_u = \mathbf{w}_u^{\hermitian} \mathbf{H}_u \mathbf{R}_{\FH}$. Similar to precoding in the SA architecture, $\kappa_{\FH}$ is the power scaling coefficient for $\|\mathbf{R}_{\FH}\mathbf{B}_{\FH}\|^2 = P_{\FH}^{(\text{out})}$ and $\alpha_{\FH}$ is the regularization coefficient.
%
%
\subsection{DAC precision}
\label{sec:DAC_noise}
The transmit noise in (\ref{eq:SINR_expression}) comes from the quantization error due to DACs with finite precision. A practical system design uses sufficient quantization precision such that the transmission noise level stays well below the receiver thermal noise. Different architectures require different values of effective number of bits (ENOB) for such goal. The required ENOB in three architectures are 
\begin{align}
\label{eq:ENOB_theory}
\tilde{B}_{\DA} =& \frac{\text{PAPR}-1.76+D+10\log_{10}\left(\frac{P^{(\text{out})}_{\DA}}{\sigma^2_{\text{n,rx}}}\right)}{6}\nonumber\\
\tilde{B}_{\SA} =& \frac{\text{PAPR}-1.76+D+10\log_{10}\left(\frac{P^{(\text{out})}_{\SA}}{\sigma^2_{\text{n,rx}}}\frac{N_{\SA}}{U^2}\right)}{6}\nonumber\\
\tilde{B}_{\FH} =& \frac{\text{PAPR}-1.76+D+10\log_{10}\left(\frac{P^{(\text{out})}_{\FH}}{\sigma^2_{\text{n,rx}}}\frac{N_{\FH}}{U}\right)}{6}
\end{align}
for transmit noise to be $D$ dB lower than AWGN. In the above equation, PAPR represents the peak to average power ratio of the input signal of each DAC. Note that these expressions are accurate when DAC quantization errors are uncorrelated, which may not be valid with small number of bits, e.g., $B=1$ bits. Derivations of (\ref{eq:ENOB_theory}) are provided in the Appendix~\ref{appendix:DAC_ENOB}.

Equation (\ref{eq:ENOB_theory}) together with (\ref{eq:maximum_gain}) indicates following facts. Firstly, with fixed signal power gain $G_{\DA}$, DACs precision in DA architecture can be reduced by increasing array size and decreasing transmit power. For SA and FH, however, the transmit noise remain constant regardless of the source of signal power gain. Secondly, with the same signal power gain and transmit power, DA architecture has lower requirement in DAC quantization as compared to SA and FH.
%
%
\subsection{Phase shifter precision}
\label{subsec:phase_shifter_resolution}
\begin{figure*}[t]
\begin{center}
\includegraphics[width=1\textwidth]{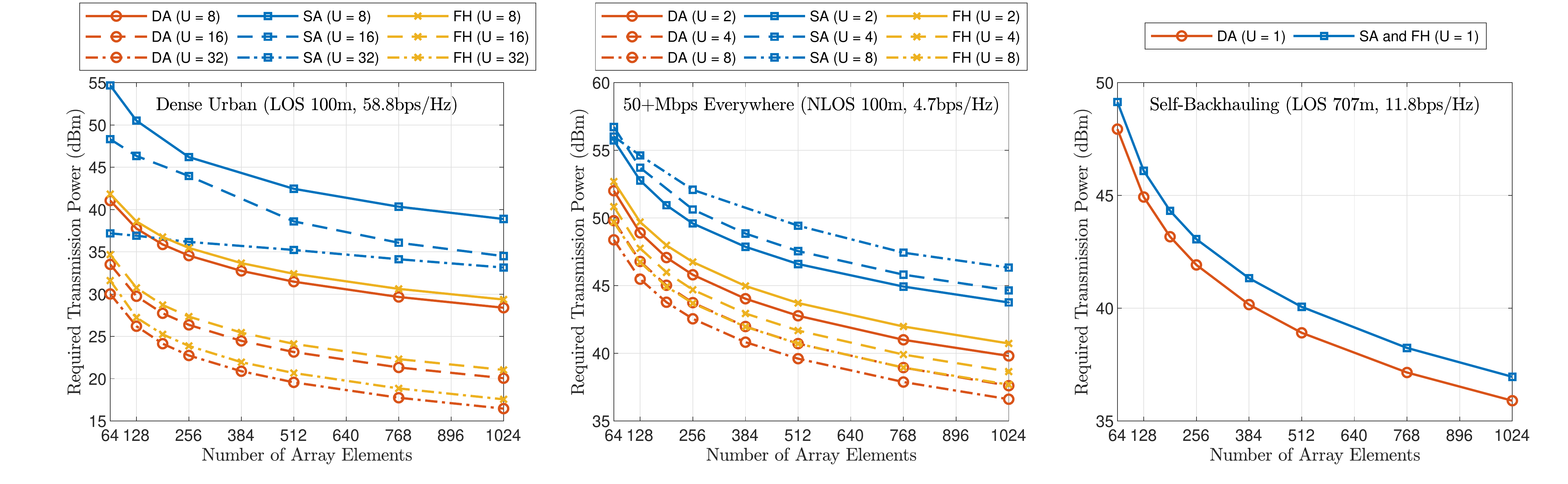}
\end{center}
\caption{The required total transmit power $P^{(\text{out})}$ with different number of array elements ($N$) to reach SE target in three typical 5G use cases.}
\label{fig:Tx_power_array_size_tradeoff}
\end{figure*}
In both SA and FH architectures, finite resolution of phase shifters leads to a changed power level of sidelobes and shifted location of nulls, as compared to system using ideal devices. More importantly, the locations of main lobe varies and associated signal gain drops. One might expect highly precise phase shifters are required to accurately control beams. In this subsection, we discuss the impact of finite resolution of phase shifters on SA and FH architectures.

The former issue regarding the distorted sidelobes is less troublesome in both SA or FH transmitter array architecture. Sidelobes lead to multi-user interference as seen from the off-diagonal elements in the effective channel $\mathbf{G}_{\text{SA}}$ and $\mathbf{G}_{\text{FH}}$. When system is aware of potential interference, digital precoding stage can be used to effectively suppress them. A practical way to acquire the information of effective channel is via a training procedure where BS and UE use quantized analog beamformer to exchange pilot symbols and estimate effective channel $\mathbf{G}_{\text{SA}}$ and $\mathbf{G}_{\text{FH}}$. This training procedure is similar to the multi-beam scheme proposed for the next generation of mmW indoor system \cite{Rice_80211ay}. Meanwhile, the gain reduction due to finite phase shifter resolution is not severe either. In fact, the gain degradation is lower bounded by 0.68dB, 0.16dB and 0.04dB with $Q = 3, 4, 5$ bits quantization of phase shifters and does not scale with the array size or multiplexing level. An analysis that supports these numbers is provided in the Appendix~\ref{appendix:phase_shifter}. Equivalently, the gain degradation is bounded by 0.16dB so long as angle error of phase shifters are no larger than 11.25 degree. Such specifications are not difficult to meet in state-of-the-art devices as it will be discussed in Section~\ref{sec:RF_power}.
%
%

\subsection{Simulation results}
\label{subsec:simulation_result}

\begin{figure*}[t]
\begin{center}
\includegraphics[width=1\textwidth]{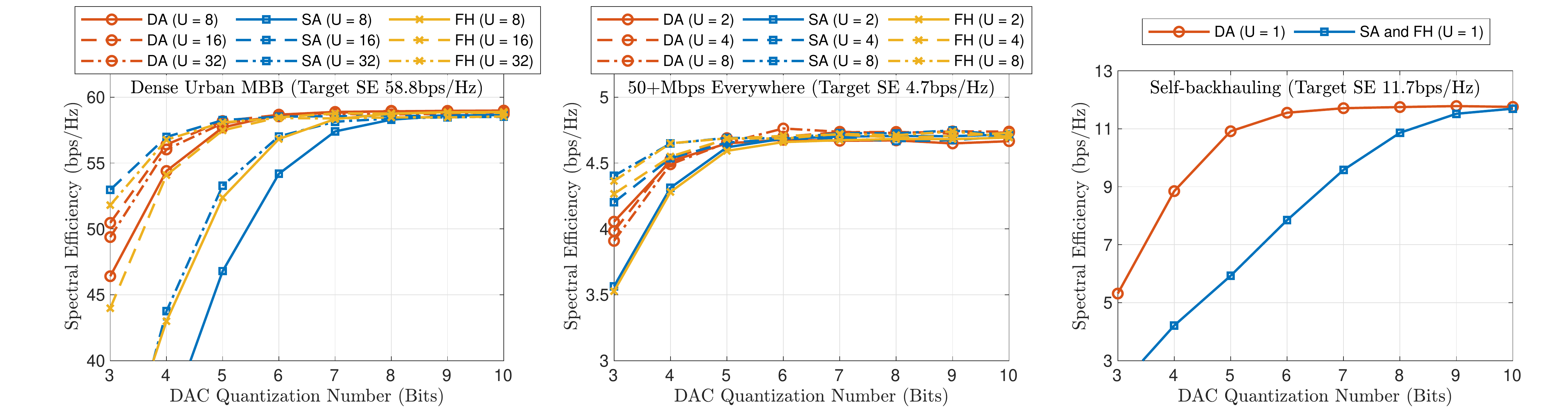}
\end{center}
\caption{SE performance with quantization on the baseband precoding and DAC. Three architectures use 256 array elements and output power is adjusted according to Figure~\ref{fig:Tx_power_array_size_tradeoff}. The baseband precoding uses fixed point operation with precision 2 bits greater than associated DAC quantization which ensures negligible degradation as compared to baseband precoding with floating point operation.}
\label{fig:DAC_quantization}
\end{figure*}
In this subsection, simulation results are presented to show the required design parameters to reach SE target in three array architectures. 

In the simulation, 3D mmW MIMO channel between BS and $U$ UEs are generated according to mmW sparse scattering model \cite{7160780}.
The channel between BS and each UE consists of 20 multi-path rays in 3 multipath cluster and LOS cluster, if exists, is 10dB stronger than the rest. Angle of arrival (AOA) and angle of departure (AOD) of clusters are uniform random variables within azimuth range $[-60^{\circ}, 60^{\circ}]$ and elevation range $[-30^{\circ}, 30^{\circ}]$. Azimuth and elevation AOA and AOD of rays within a cluster have random deviations from the cluster specific AOA and AOD, and they follow zero mean Laplacian distribution with $10^{\circ}$ standard deviation. In dense urban MBB, a scheduler is assumed such that the LOS paths of all target receivers are unique \cite{7888765}. The mean SE is evaluated by taking average of SINR in (\ref{eq:SINR_expression}) over $U$ UEs and use Shannon capacity formula, i.e., $\text{SE} = \sum_{u=1}^{U}\log_2(1+\text{SINR}_u)$. The data streams used in the simulation are Gaussian distributed and their magnitudes are truncated such that PAPR is 10dB. 

With ideal hardware, the required transmit power $P^{(\text{out})}$ to reach SE target with various antenna size $N$ and number of data streams $U$ in three architectures are shown in Figure~\ref{fig:Tx_power_array_size_tradeoff}. 

We first focus on how transmit power changes with parameter $N$ and $U$. Increasing array size $N$ is effective in reducing transmit power in all scenarios since it helps improve both signal gain and interference control from narrow beams. When interference from multi-beam is negligible, the transmit power saving from increasing $U$ depends on difference between the SINR target reduction in Table~\ref{tab:link_budget_use_cases} and signal gain dropping in (\ref{eq:maximum_gain}). 
For example, when $U$ increases from 2 to 4 and 4 to 8 in MBB, the SINR requirement reduces by 5.2dB and 4dB. Meanwhile, the signal gain changes by 3dB, 6dB and 3dB in DA, SA, and FH, respectively. Therefore DA and FH save around 2.2dB and 1dB $P^{(\text{out})}$ and SA is forced to use around 0.8dB and 2dB higher $P^{(\text{out})}$. 
It is also true in high-$N$ regime of DA and FH in the Dense Urban MBB. When $U$ increases from 8 to 16 and 16 to 32, the SINR requirement reduces by 11.4dB and 6.6dB. Therefore the power saving at $N=1024$ is around 8.4dB and 3.3dB for both DA and FH. Power saving is more difficult to predict when system needs to trade power gain for interference control. Therefore the transmit power saving from increasing $U$ with smaller antenna $N$ and large multiplexing $U$ is less accurately using the above analysis.

Then we focus on the comparison between array architectures. There is one universal conclusion that holds true for DA and FH in all scenarios. DA and FH have the same maximum signal gain when $P^{(\text{out})}$ and $N$ are the same according to (\ref{eq:maximum_gain}). In simulation, FH actually requires near 1dB higher $P^{(\text{out})}$ than DA in all scenarios. This gap is due to the loss from the two-stage precoding of FH. Further exploiting hardware capability, e.g., using phase-and-magnitude analog precoders, and designing better hybrid precoding algorithm in FH would reduce this gap.

Next, we compare array architectures in each use case. In self-backhauling where data stream number $U$ is constraint by point-to-point environment, SA has the same performance as FH as both architectures become the same in model (\ref{eq:received_signal_model}). They both require 1dB higher transmit power than DA.
Secondly, the difference of required transmit power between architecture can by analyzed by (\ref{eq:maximum_gain}) in 50+Mbps Everywhere. Equation (\ref{eq:maximum_gain}) reveals that SA has $U$ times lower power gain than other architectures and it is shown in the figure that that SA requires $U$ times higher $P^{(\text{out})}$ than FH for the same performance. Equation (\ref{eq:maximum_gain}) predicts the gap between curves well in the since there is negligible interference with small number of beams.
Thirdly, in MBB use case the required transmit power gap between SA and FH in Dense Urban MBB meet (\ref{eq:maximum_gain}) when $N$ is large, i.e., SA requires to use $9, 12, 15$dB higher $P^{(\text{out})}$ than FH when $U=8,16,32$ beams are used. However, the transmit power gap between SA and FH deviates from what (\ref{eq:maximum_gain}) predicts when $N$ is small. This deviation is due to power gain and interference control trade-off. Dense Urban MBB features a large number of simultaneous data streams and the mutual interference among streams becomes system bottleneck when beam-width is not small enough.  
%
%
With $U=8$, the transmit power gap between SA and FH increases from 9dB to 13dB when $N$ reduces from $1024$ to $64$. The additional 4dB gap is the cost of controlling interference in SA, because the SA uses nearly $U$ times wider beam to carry each data stream as compared to FH. Further, the BB precoding of SA is forced to sacrifice more gain for interference control. 
With $U=32$, the gap reduces from 9dB to 6dB when $N$ reduces from $1024$ to $64$. One may expect each data stream in SA is carried by wide beams with $N/U=2$ antennas and conclude the opposite results. However, with $U=32$ data streams, each RF-chain is connected with at most $N/U=2$ antennas and such architecture is effectively a digital array. In fact, the BB precoding stage in SA facilitates each stream to be transmitted by nearly all antenna elements and improves the signal gain. In fact, the intuition of hybrid precoding approach \cite{7160780} may not be true and a better hybrid precoding scheme tailored for this regime would provide more additional power saving for SA.

\begin{figure}[t]
\begin{center}
\includegraphics[width=0.5\textwidth]{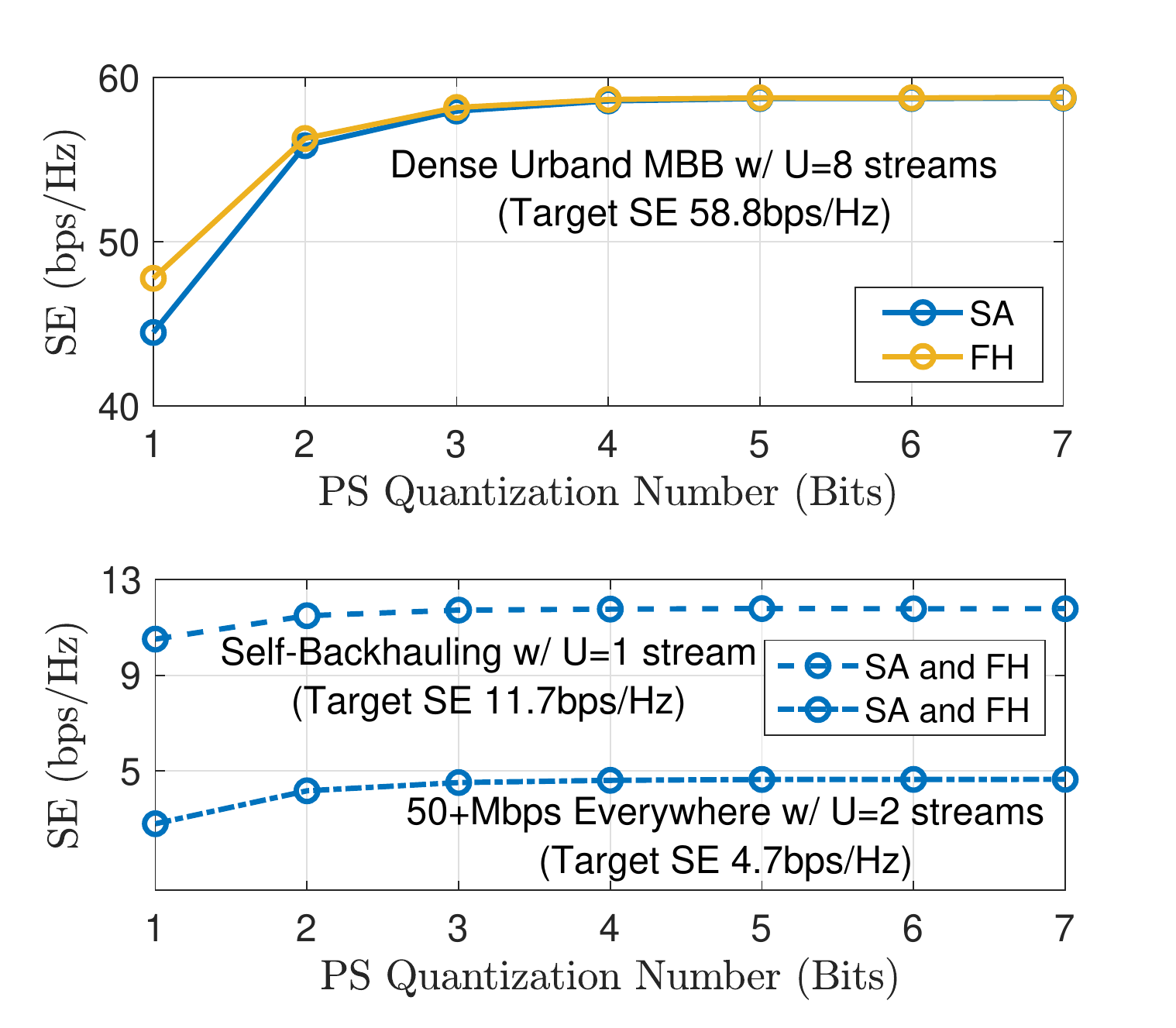}
\end{center}
\caption{SE performance with quantization on RF phase shifter of sub-array and fully-connected hybrid array. Both architectures use 256 array elements without baseband and DAC quantization error and output power is adjusted according to Figure~\ref{fig:Tx_power_array_size_tradeoff}. for target SE.}
\label{fig:PS_quantization}
\end{figure}

With finite precision in the baseband precoding, DAC and phase shifters, the SE performance is shown in Figure~\ref{fig:DAC_quantization} and Figure~\ref{fig:PS_quantization}. For clarity, all array architectures use 256 antenna elements and the transmit power $P^{(\text{out})}$ in each architecture is chosen such that it delivers the same SE performance as in quantization free cases. Figure~\ref{fig:DAC_quantization} shows the required quantization bits in baseband precoding and DAC and it matches with the analysis. According to (\ref{eq:ENOB_theory}), the required ENOB for transmit noise to be $D = 15$dB lower than AWGN in the Dense Urban MBB with $U=8$ streams are 5.1, 8.0, and 7.7 in DA, SA, and FH architectures, respectively. The SE improvement in Figure.~\ref{fig:DAC_quantization} is saturated once DAC quantization bits are beyond these values. Equation (\ref{eq:ENOB_theory}) also precisely matches with Self-backhauling use case where DA, SA, and FH requires 5.8, 10.0, and 10.0 ENOB, respectively. It is worth noting that the additive quantization error model becomes inaccuracy when the analytical ENOB from (\ref{eq:ENOB_theory}) is significantly small. For example, equation (\ref{eq:ENOB_theory}) estimates that system requires 1 to 4bits for the most scenarios in 50+Mbps Everywhere, while the required ENOB from simulation is close to 5bits. A rule-of-thumb is to use at least 5 bits. Note that this inaccuracy regime of (\ref{eq:ENOB_theory}) does not affect power consumption estimation of the system, because the direct current (DC) power of DAC does not effectively reduce by using less than 5 bits due to the fixed hardware overhead and it is discussed in details in Section.~\ref{sec:DSP_SERDES_DAC_power}.

Moreover, the precision requirement in baseband precoding and DAC of DA is in general lower than hybrid architectures throughout all scenarios and it suggests a system level power consumption saving. Last, Figure~\ref{fig:PS_quantization} shows that with the hybrid precoding approach in Section~\ref{sec:precoder_design}, the SE performance is negligibly affected by phase shifter quantization and it matches with our analysis in Section~\ref{subsec:phase_shifter_resolution}.

In summary, for the same target SE performance, DA requires a reduced transmit power or number of array elements as compared to SA and FH. Besides, the DAC quality of DA is relaxed as compared to the hybrid architecture. A fair comparison among architectures cannot overlook these factors by restricting architectures to use the same transmit power, number of array elements, or specification of hardware components. The design parameter trade-off analyzed in this section leads to a more practical comparison in Section~\ref{sec:results}.

\section{Hardware Design Challenges of Transmitter Array}
\label{sec:IC_PCB_implementation_discussion}

In this section, we discuss practical hardware design of mmW arrays with different architectures. We first introduce the distributed array processor module. Then, the necessary circuits blocks for baseband signal and RF signals distribution are discussed.

\begin{figure*}[t]
\begin{center}
\includegraphics[width=1\textwidth]{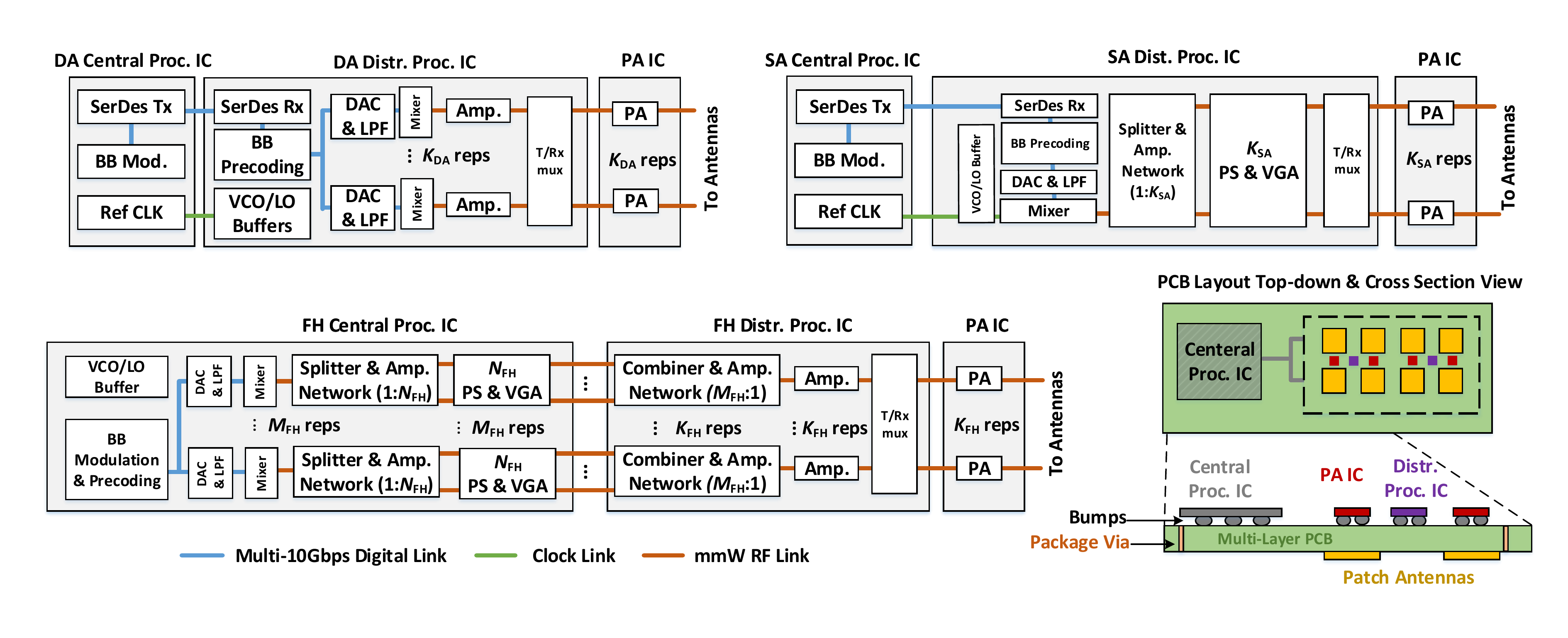}
\end{center}
\caption{The proposed mmW antenna array circuits layout plan of three architectures. The functionality of tranceivers arrays are divided into a centralized processing IC (Central Proc. IC), multiple distributed processing ICs (Distr. Proc. IC) and PA IC which are close to patch antenna elements to reduce mmW signal Loss. Other abbreviations used in the figure include baseband modulation (BB Mod.), anti-aliasing low-pass filters (LPF), transmitter/receiver multiplexer (mux), reference clock (Ref CLK), amplifiers (Amp.), and repetitions of elements (reps).}
\label{fig:implementation_plan}
\end{figure*}
%
%
\subsection{Distributed array module}
\label{subsec:distribution_array_module}
The conventional MIMO system integrates array processing module in an IC and delivers RF signal to antennas. Such centralized design may not be practical in mmW system with massive number of antennas. With a compact and centralized IC, mmW signals routed to hundreds of array elements suffer severe insertion loss\footnote{The wavelength at 28GHz band is 10.7mm, 256 antennas in a square alignment with half-wavelength require at least 7327$\text{mm}^2$.}. Besides, the heat dissipation becomes a concerns for a centralized solution. Moreover, array size scalability becomes challenging since adding more elements requires completely new processing module. 

A practical solution is to implemented processing hardware for antenna arrays in a distributed manner \cite{7355304}. In DA and SA, each IC in a processing module integrates the processing circuits for $K_{\DA}$ and $K_{\SA}$ antennas and is located close to these antennas. Although a centralized digital processor is still necessary for some baseband functionality, e.g., symbol mapping and channel coding, the digital baseband precoding can be implemented in each distributed module. With such design, the system needs to deliver $U$ digital signal streams rather than $M$ digitally precoded signal streams to the processing modules \cite{7355304}. It offers a significant saving of baseband signal distribution throughput given $M\gg U$ in DA and SA. The DAC, upconverter and RF signal processing are also included in the processing module.  The digital signals from central processor are routed and recovered through Serializer/Deserializer (SerDes) sub-system in each of the processing modules. Note that the exact value of elements integrated in an IC affects system area and energy. But the discussion of that is beyond the scope of this work. The patch antenna is directly attached on the printed circuits board (PCB).

The illustration of distributed DA hardware implementation is shown in Figure~\ref{fig:implementation_plan}. In the remaining of the paper, power consumption and cost estimation of DA system is based on design where each module contains $K_{\DA} = 8$ antenna elements and associated processing circuits. Each DA module contains SerDes, voltage controlled oscillator (VCO) within a phase-lock-loop (PLL), and RF-chains and T/Rx multiplexers. The power amplifiers for 5G mmW applications are expected to be built in non-silicon material, as shown in Section~\ref{sec:RF_amplification_power} and they are placed next of DA processing IC.






The illustration of SA implementation is illustrated in Figure~\ref{fig:implementation_plan}. In SA, each module has processing circuits for $K_{\SA}$ antenna elements. Each of them contains SerDes, VCO and phase shifter networks. 


There is no priori work on FH implementation with larger than 8 antennas. The RF signal routing is a challenging task in FH architecture, because the input signal for each antenna element is a combination of signals from all RF-chains. The most viable approach we could anticipate is illustrated in Figure~\ref{fig:implementation_plan}. Opposite of DA and SA architectures, routing loss cannot be reduced by distributing RF-chains into a closer position, since their outputs are required to be delivered to entire PCB board. In the proposed design, each array module integrates a combining network and delivers the combined signal to nearby antenna elements. It also contains RF amplifiers to compensate for insertion loss during the RF signal routing and combining. 

In all array architectures, routing digital baseband signal and RF signals plays a critical roles. We discuss associated challenge and solutions in the next subsections.


%
%
\subsection{BB signal distribution}
The digitally precoded sample streams require to be routed into each processing module by serial-link tranceivers in all array architectures. The state-of-the-art SerDes supports data rates over 50Gb/s using PAM-4 signaling in wireline chip-to-chip communication. The specific design of SerDes system is beyond the scope of this work. In Section ~\ref{sec:power_cost_modeling}, we use the specifications of ultra-high-speed tranceivers.



%
%
\subsection{RF signal distribution}
Multiple circuit components introduce non-negligible insertion losses that need to be carefully handled by system designers.
\begin{itemize}
  \item \textit{PCB and Inter-Connectors Loss:}
  RF signal suffers from interconnect loss between the silicon chip RF ports and the antenna elements.  The low-loss PCB board, such as RO 3000 series and 4000 series, 28GHz signal have 1.25dB/inch insertion. Besides, each IC chip require to be placed on organic or ceramic substrate (interposer) to distribute the chip ports to a ball-grid array and it has an additional 1-2dB distribution loss. This implementation loss needs to be pre-compensated before the RF signal is fed into antenna.
  
  \item \textit{Intra-Chip Transmission Lines Losses:}
  RF signal loss in silicon is significant at mmW band. According to \cite{7337695}, there is up to 0.6dB/mm transmission line loss at 28GHz. The length of transmission line is proportional to the IC size but exact value is determined by actual IC design. According to a 60GHz array design \cite{7747488}, phase shifter and Wilkinson RF splitter take most of the IC area. The intra-chip routing loss can be roughly estimated by taking into account the required area of those components. With the practical components size in Section~\ref{sec:power_cost_modeling}, the loss in an SA module with $K_{\SA} = 32$ phase shifters is less than 1dB but up to 3-4dB for FH since each RF-chain distributes signals into hundreds of phase shifters that require dozens of millimeters square area.

  \item \textit{Power Splitters and Combiners Loss:}
  In the analog beamforming stage of SA and FH architectures, output signals of RF-chains need to be fed into phase shifter network for phase rotation. The Wilkinson power splitters are commonly used for such purpose \cite{7747488,7969017,7969030}. Moreover, the fully-connected hybrid architecture uses same Wilkinson structure to combine multiple RF signals before power amplification. An ideal power splitter/combiner introduces 3dB insertion loss in each of the one-to-two splitter (1:2) or two-to-one combiner (2:1) unit. Practical design often has an additional 1dB implementation loss. It results in a $4\log_{2}(K_{\SA})$ dB power drop in the SA architecture. For FH architecture, the splitters and combiners introduce total $4\log_{2}(N_{\FH}M_{\FH})$ dB loss. 
\end{itemize}
All the above RF insertion losses lead to an reduced EIRP at the antenna and therefore need to be properly compensated. The detailed distribution budget in all architectures is discussed in Section.~\ref{sec:RF_amplification_power}.
%
%
\section{Hardware Power and Cost Modeling}
\label{sec:power_cost_modeling}
In this section, we first provide the power and cost model of necessary circuits blocks based on a survey of the state-of-the-art circuits design and measurement. The power consumption contains DSP module for precoding, SerDes, mixed signal components, and RF components. Note that other hardware blocks such as power supply, active cooling may consume considerable power \cite{6056691}. We omit them in this work since these are constant hardware overhead. Then, examples are provided for signal distribution budgets calculation in order to determine necessary RF amplifiers to compensate insertion loss. Finally, we summarize the total power and cost calculating formula for all architectures operating with different design parameters.

\subsection{Digital signal processing power}
\label{sec:DSP_SERDES_DAC_power}
Due to large bandwidth, the array processing in the digital baseband needs to support such high throughput. The DSP for array processing mainly consumes power for digital precoding and digital signal routing. Note that tasks such as channel coding, higher layer processing in the communication standard stack are not included since they have equal power consumption for all architectures. Channel estimation and precoder computation are also omitted since they occur at time scale that is several orders of magnitude longer than symbol duration. 

The DSP power estimation contains linear precoding and 4096 point inverse discrete Fourier transform\footnote{We assume $N_{\text{FFT}} = 4096$ point IDFT for 850MHz signal bandwidth to achieve 3GPP-specified subcarrier spacing 240KHz \cite{5GNR_rel15}} (FFT). The precoding requires multiplication of $M \times U$ complex matrix with $U\times 1$ complex vector. It has $6UM$ fixed points operations. Note that the number of operation does not change with different design choices of $N_{\text{FFT}}$, because the number of precoder slices in sub-carriers and symbol duration change. The latter consists of $\log_2(N_{\text{FFT}}) = 12$ complex multiplication per sample per RF-chain, and it results in $6\times12M \times \text{BW}$ operations per second. We use $\text{FOM}_{\text{DSP}}  = 13\text{GOPS/mW}$ in 40nm CMOS as state-of-the-art fixed point digital computation efficiency \cite{6858388}. As a consequence, the power consumption in the digital precoding is
\begin{align}
P_{\text{Precoding}} = \frac{(6UM+72M ) \times  \text{BW}}{\text{FOM}_{\text{DSP}}} 
\label{eq:power_precoding}
\end{align}
where BW is the signal bandwidth. The power consumption $P_{\text{Precoding}}$ has unit Watt. 

The power of SerDes system is modeled in the following equation

\begin{align}
P_{\text{SerDes}} = \text{FOM}_{\text{SerDes}} \times \text{BW}_{\text{OS}}\times \text{ENOB} \times  U 
\label{eq:power_SERDES}
\end{align}
In the above, $\text{ENOB}$ is the required precision in the digital precoding and DAC of mmW transmitter and its value is determined according to the analysis in (\ref{eq:ENOB_theory}) and Figure~\ref{fig:DAC_quantization}. $P_{\text{SerDes}}$ scales with the number of independent data stream $U$ due to the distributed digital precoding. The figure-of-merit of SerDes is adopted as $\text{FOM}_{\text{SerDes}} = 10\text{mW/(Gb/s)}$ \cite{7417905} in this work. Note here we use $\text{BW}_{\text{OS}}$ as the oversampled data rate after considering a factor of 2 oversampling ratio, i.e., $\text{BW}_{\text{OS}} = 1.7$GS/s.

%
%
\subsection{Power model of mixed signal components}
In section~\ref{sec:DAC_noise}, we analyze the impact of DAC quantization in different array architecture. The DAC power consumption is mainly determined by the sampling frequency and effective number of bits. The total power consumption in each DAC is computed using the following equation
\begin{align}
P_{\text{DAC}} = \text{FOM}_{\text{DAC}} \times  \left(2^{\text{ENOB}} \times \text{BW}_{\text{OS}}\right) + P_{\text{buffer}} 
\label{eq:power_DAC}
\end{align}
where $P_{\text{DAC}}$ has unit. $\text{BW}_{\text{OS}}$ and are similarly define in (\ref{eq:power_SERDES}). The state-of-the-art specification of DAC is $\text{FOM}_{\text{DAC}} = 0.08$PJ/conversion \cite{7062924}. A constant hardware overhead for signal amplification is modeled as $P_{\text{buffer}}$ = 10\text{mW} for $-14$dBm output signal power. Therefore further reducing precision has limited power saving benefits when $P_{\text{buffer}}$ dominates.

%
%
\subsection{Power model of RF signal components}
\label{sec:RF_power}
In this section, we estimate the required power consumption in the RFIC, including the power for signal amplification and analog array processing for hybrid architecture. The components are phase shifter, local-oscillator using phase-lock-loop (PLL), mixer, RF amplifier for gain compensation, and the power amplifier for transmission.
\begin{itemize}
  \item \textit{Local oscillator (LO) and mixer:} 
  The phase noise of an oscillator is inversely proportional to the power dissipated \cite{7355304}. The state-of-the-art VCO design \cite{7062991,8094573,7508265,6720214} facilitates phase noise lower than -110dBc/Hz at 1MHz by using less than 30mW DC power consumption, and system performance is not affected by such noise specification \cite{7116610}. Considering the required buffer at the output, the power consumption of VCO block can be $P_{\text{VCO}} = 60\text{mW}$ for each element. Mixer can be made by active or passive devices. Practically, passive mixers are easier to implement and have better linearity and noise. Mixers require enough LO signal power to be driven. In this work, we select the input LO power to be at least -5dBm and the power consumption of mixer is $P_{\text{Mixer}} = 10\text{mW}$. The total power consumption of LO is $P_{\text{LO}} = 70\text{mW}$

  \item \textit{Phase shifter:} RF phase shifting can be implemented in various ways, see \cite{poon2012} for a comprehensive survey. The state-of-the-art work uses reflective-type phase shifter (RTPS) and switch-type phase shifter (STPS) as main approaches of passive PS \cite{7956180,7969012,7969017,7358176,7268915}. Such approaches use delay line with controllable length to generate desired phase shifting. Although nearly zero DC power consumption is required, passive PS often has high insertion loss and large IC area due to the delay line. The active approach uses vector modulator (VM), which consists of variable gain amplifier in both In-Phase and Quadratic RF path to generate a complex gain as magnitude adjustment and phase shifting coefficient. VM requires active devices and has higher power consumption than STPS or RTPS. Meanwhile, VM requires less IC area \cite{7969030,7747488,6259477}. In this work, we use VM for building block of hybrid architecture and the power model is $P_{\text{PS}} = 10\text{mW}$ with $2$dB gain.

\end{itemize}

%
%
\subsection{RF signal amplification power}
\label{sec:RF_amplification_power}
\begin{figure}
\begin{center}
\includegraphics[width=0.5\textwidth]{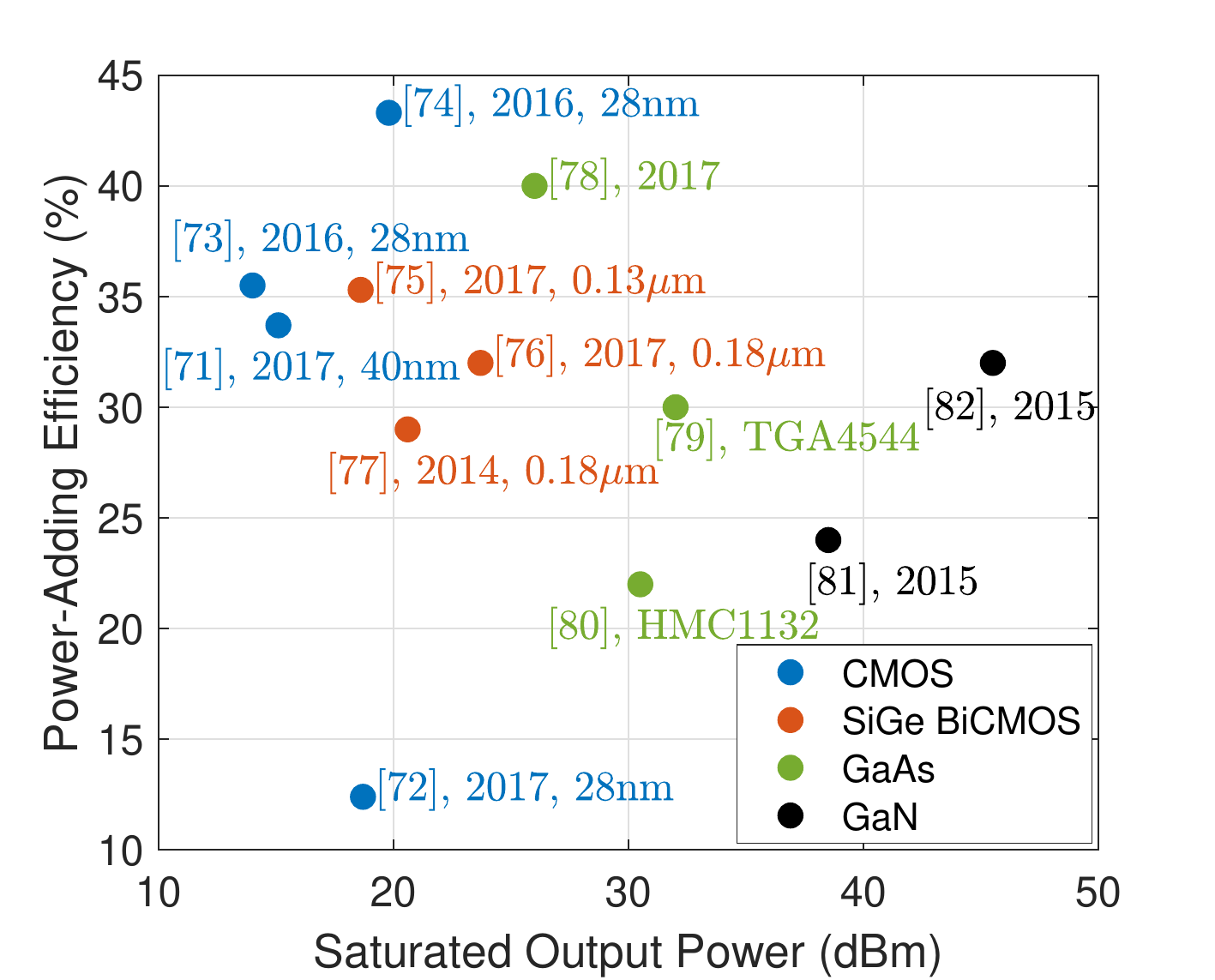}
\end{center}
\caption{The saturated output power and power-added efficiency at that power level ($\text{PAE}_{\text{max}}$) of the state-of-the-art power amplifiers at 28GHz. Data are from \cite{PA_CMOS_2017_Shakib,PA_CMOS_2017_Moret, PA_CMOS_Shakib_JSSC_2016, PA_CMOS_2016_Park, PA_BiCMOS_2017_Sarkar,PA_BiCMOS_2017_JSSC_Sarkar, PA_BiCMOS_2014_Kim , PA_GaAs_2017_IMS_Nguyen, qorvo_TGA4544, AD_HMC1132 , PA_GaN_2015_Fujii,PA_GaN_2015_Din} and labels include publication date and silicon technology.}
\label{fig:PA_specs}
\end{figure}

\begin{figure*}[t]
\begin{center}
\includegraphics[width=1\textwidth]{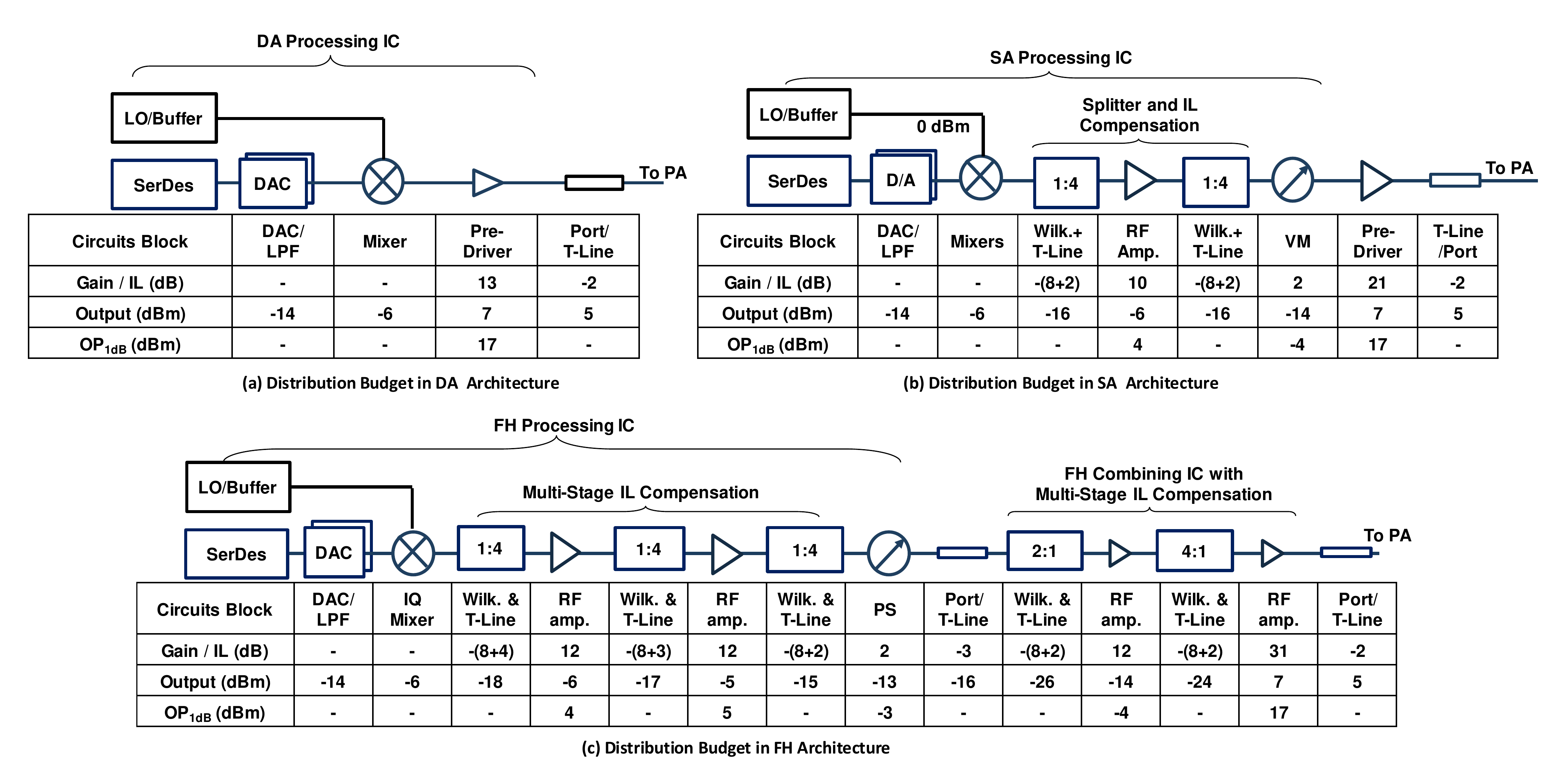}
\end{center}
\caption{The signal distribution budget example of three array architectures.}
\label{fig:distribution_budget}
\end{figure*}

The RF signals amplification has two categories: gain compensation amplifier and power amplifier.
\begin{itemize}
  \item \textit{RF amplifier:}
Gain compensation amplifiers are used to compensate insertion loss in the analog beamforming for hybrid architectures. As discussed in Section~\ref{sec:IC_PCB_implementation_discussion}, hybrid architectures require to distribute up-converted RF signal into phase shifter networks. During this procedure, insertion loss is introduced in power splitter, transmission line and power combiner. These losses need to be properly compensated in order to deliver sufficient radiated signal power at the antenna. From the cost perspective, it is better to provide the gain before power splitting occurs since it requires fewer number of amplifiers. However, it raises the linearity concern of CMOS amplifier. As it is shown in the next subsection, a large hybrid array has more than 20dB insertion loss in the distribution route and in order to pre-compensate such loss immediately after up-conversion leads to a severe nonlinear distortion in RF signals. A practical design typically places amplifiers in a hierarchical manner along RF signal distribution route \cite{7747488}. Besides, the gain compensation amplifiers need to be carefully designed and their power consumption cannot be overlooked. The power model adopted in this work considers gain compensation amplifier design from \cite{7337695}, where each amplifier has up to 15dB gain with $P_{\text{Amp}} = 40\text{mW}$ power consumption. Note that active combining \cite{7969030} is an alternative approach that combines RF signal in current mode using low-noise amplifiers. Although insertion loss can be avoided, there is power consumption in each combiner. We do not discuss this approach in details.

\begin{table*}[t]
\caption{Summary of circuits blocks in array architectures}
\centering
\begin{tabular}{c|c|c|c|c|c|c|c|c|c}
\hline
\hline
  Circuits Block & BB DSP & $\text{SerDes}^{\text{h}}$ & DAC & LO/Mixer & PS & $\text{Splitter/Combiner}^{\text{b}}$ & RF $\text{Amp.}^{\text{}}$ &  RF Amp. & PA \tabularnewline
  & &  &  &  &  & & (IL Comp.) &  (Pre-Driver) & \tabularnewline
\hline
\hline
DC Power per Block& eq.(\ref{eq:power_precoding}) & eq.(\ref{eq:power_SERDES}) & eq.(\ref{eq:power_DAC}) & 60+10mW & 10mW & - & 40mW & 40mW &  eq.(\ref{eq:power_PA}) \tabularnewline
\hline
IC Area per Block ($\text{mm}^2$) & Varies$^{i}$ & $1.21^{\text{c}}$ & $0.05^{\text{d}}$ & $0.18^{\text{e}}$ & $0.05^{\text{f}}$ & $0.04^{\text{g}}$ & $0.025^{\text{f}}$ & $0.025^{\text{f}}$  & - \tabularnewline
\hline
\hline
Blocks in $\text{DA}^{\text{a}}$ & 1 & $N_{\DA}/K_{\DA}$&\multicolumn{2}{c|}{$N_{\DA}$} & - & $N_{\DA}$ & - &  $N_{\DA}$ & $N_{\DA}$ \tabularnewline
\hline
Blocks in $\text{SA}^{\text{a}}$ &  1 &$N_{\SA}/K_{\SA}$ &\multicolumn{2}{c|}{$N_{\SA}/K_{\SA}$} & $N_{\SA}$ & $N_{\SA}-N_{\SA}/K_{\SA}$ & $N_{\SA}/4$ & $N_{\SA}$ & $N_{\SA}$ \tabularnewline
\hline
Blocks in FH &  1 & - &\multicolumn{2}{c|}{$U$} & $UN_{\FH}$ & $2N_{\FH}U-N_{\FH}-U$ & $2UN_{\FH}/3$ & $N_{\FH}$ & $N_{\FH}$ \tabularnewline
\hline
\hline
Blocks per DA Antenna & N/A & $1/N_{\DA}$&\multicolumn{2}{c|}{1} & - & 1 & - & 1  & 1 \tabularnewline
\hline
Blocks per $\text{SA}^{\text{a}}$  Antenna &  N/A & $1/K_{\SA}$ &\multicolumn{2}{c|}{$1/K_{\SA}$} & 1 & $1-1/K_{\SA}$ & 1/4 & 1 & 1 \tabularnewline
\hline
Blocks per FH Antenna &  N/A & - & \multicolumn{2}{c|}{$U/N_{\FH}$} & $U$ & $2U-1-U/N_{\FH}$ & $2U/3$ & 1 & 1 \tabularnewline
\hline
\multicolumn{10}{l}{a. We do not focus on varying the number of elements in module. $K_{\DA} = 8$ and $K_{\SA} = 16$ are treated as constants.}\tabularnewline
\multicolumn{10}{l}{b. It refers to a 1:2 or 2:1 Wilkinson splitting or combining unit.}\tabularnewline
\multicolumn{10}{l}{c. We use 0.89$\text{mm}^2$ \cite{7417905} and 0.32$\text{mm}^2$ \cite{7417910} for SerDes receiver and transmitter respectively. They are fabricated in 28nm and 16nm CMOS.}\tabularnewline
\multicolumn{10}{l}{d. Specification is from \cite{7062924} and the DAC has 8 bits precision and uses 28nm fabrication.}\tabularnewline
\multicolumn{10}{l}{e. Specification of 28GHz LO and mixer are from \cite{7508265} and 65nm CMOS fabrication is used.}\tabularnewline
\multicolumn{10}{l}{f. Specification is estimated from figure in \cite{7747488}. 0.18$\mu$m BiCMOS is used for fabrication.}\tabularnewline
\multicolumn{10}{l}{g. Specification is estimated from figure in \cite{7747488} and scaled by wave-length due to its direct impact in Wilkison divider.}\tabularnewline
\multicolumn{10}{l}{h. Assuming SerDes module is used for each module define in Section.~\ref{subsec:distribution_array_module}.}\tabularnewline
\multicolumn{10}{l}{i. $A_{\text{Precoding}} = \frac{(6UM+72M ) \times  \text{BW}}{\text{FOM}_{\text{DSP,area}}} $ with $\text{FOM}_{\text{DSP,area}} = 500$GOPS/$\text{mm}^2$, 10 times scaling from \cite{6982237} due to potential advanced CMOS process.}\tabularnewline
\end{tabular}
\label{tab:HW_specs_summary}
\end{table*}

\item \textit{Power amplifier (PA):}
Power amplifiers consume large amount of power in current base-stations operating in sub-6GHz band. In the mmW BS system design there are two conflicting scaling direction. On one hand, the transmit power of each PA is relaxed due to the use of massive antenna array for similar total power. On the other hand, the power amplifier efficiency is lower than those designed for sub-6GHz band. In Figure~\ref{fig:PA_specs}, specifications of the state-of-the-art mmW power amplifier at 28GHz are shown. Specifically, the power-added-efficiency (PAE) at saturated output power and associated saturated output power are presented. Different semiconductor technologies, e.g., CMOS, BiCMOS, Gallium Arsenide (GaAS), and Gallium Nitride (GaN) are included. The state-of-the-art CMOS or SiGe BiCMOS PAs are not suitable due to the low saturated output power. Assuming 10dB PAPR margin, even with an extremely large array of 1024 elements, the 46dBm total transmitter power leads to 16 dBm output power for each element. Thus the PA is likely to require a saturation point of 26dBm and this is a challenging target for PAs suitable for deployment in arrays. GaAs PAs are generally cheaper than GaN PAs and are expected for 5G array applications without operating in strongly nonlinear region. In the proposed PA power consumption model, a PA efficiency is $\eta_{\text{PA}} = 0.185$ is adopted. Specifically, the calculation of PA efficiency is based on 0.3 peak PAE, 10dB power back-off, and a Doherty PA architecture\footnote{In Doherty PA, the PAE remain constant when the instantaneous output magnitude $a$ is no more than 3dB weaker than the peak magnitude $a_{\text{max}}$, i.e., $\text{PAE}(a) = \text{PAE}_{\text{max}}, a\geq a_{\text{max}}/2$. Otherwise, the PAE drops as a linear function of instantaneous output magnitude, i.e., $\text{PAE}(a) = \frac{2a}{a_{\text{max}}}\text{PAE}_{\text{max}}, a< a_{\text{max}}/2$. Thus, the average efficiency is $\eta_{\text{PA}} = \int_{a}f_{A}(a)\text{PAE}(a)da$, where $f_{A}(a)$ is the probability distribution of signal magnitude. When $\text{PAE}_{\text{max}}=0.3$ and the signal magnitude is Rayleigh distributed with average power 10dB below the peak, PA efficiency is $\eta_{\text{PA}} = 0.185$.}. Accordingly, the power consumption in each PA element is
\begin{align}
P_{\text{PA}} = \frac{P^{(\text{out})}}{N\eta_{\text{PA}}},
\label{eq:power_PA}
\end{align}
where the number of array elements $N$ and output power $P^{(\text{out})}$ are from Figure~\ref{fig:Tx_power_array_size_tradeoff} in each architecture.
\end{itemize}

%
%
\subsection{Summary of specifications of circuits blocks for transmitter array architectures}
\label{subsec:distribution_loss_summary}
In Figure~\ref{fig:distribution_budget}, we present the signal distribution budget example of three array architectures with 64 elements. Specifically, we focus on the insertion loss in PCB, silicon, and RF devices as modeled as in Section~\ref{sec:IC_PCB_implementation_discussion}. There is more than 10dB loss for every two stages of Wilkinson splitters/combiners plus associated transmission line. As a consequence, RF amplifiers can be placed to compensate such loss in SA as shown in Figure~\ref{fig:distribution_budget}(b). For FH, multi-stage compensation is required to avoid saturation as shown in Figure~\ref{fig:distribution_budget}(c). Such design is commonly adopted in implementation of phased array \cite{7747488}. Moreover, a combining network in FH also needs similar design. For a splitting or combining network with $N_{\text{wilk}}$ ports, we use an approximation number of $\sum_{n=1}^{\infty}N_{\text{wilk}}/4^n \approx N_{\text{wilk}}/3$ amplifiers for simplicity. Therefore, FH requires a total $UN_{\FH}/3$ amplifiers in both splitting and combining network. Moreover, for all architectures, we assume a 5dBm signal strength is required at the input of PA \cite{AD_HMC1132,qorvo_TGA4544}. The output of each mixer is -6dBm. For a Wilkinson splitter or combiner with $N_{\text{wilk}}$ ports, a total $N_{\text{wilk}}-1$ splitting (1:2) or combining (2:1) units are required. As a consequence, the required number of Wilkinson units are $(K_{\SA}-1)N_{\SA}/K_{\SA}$ and $U(N_{\FH}-1)+N_{\FH}(U-1)$ in the SA and FH architectures, respectively. A summary of specifications of circuits blocks, total number of blocks in each architectures, and required number of blocks per antenna element are summarized in Table~\ref{tab:HW_specs_summary}.






%
%
\section{Comparison results}
\label{sec:results}
\begin{figure*}[t]
\begin{center}
\includegraphics[width=1\textwidth]{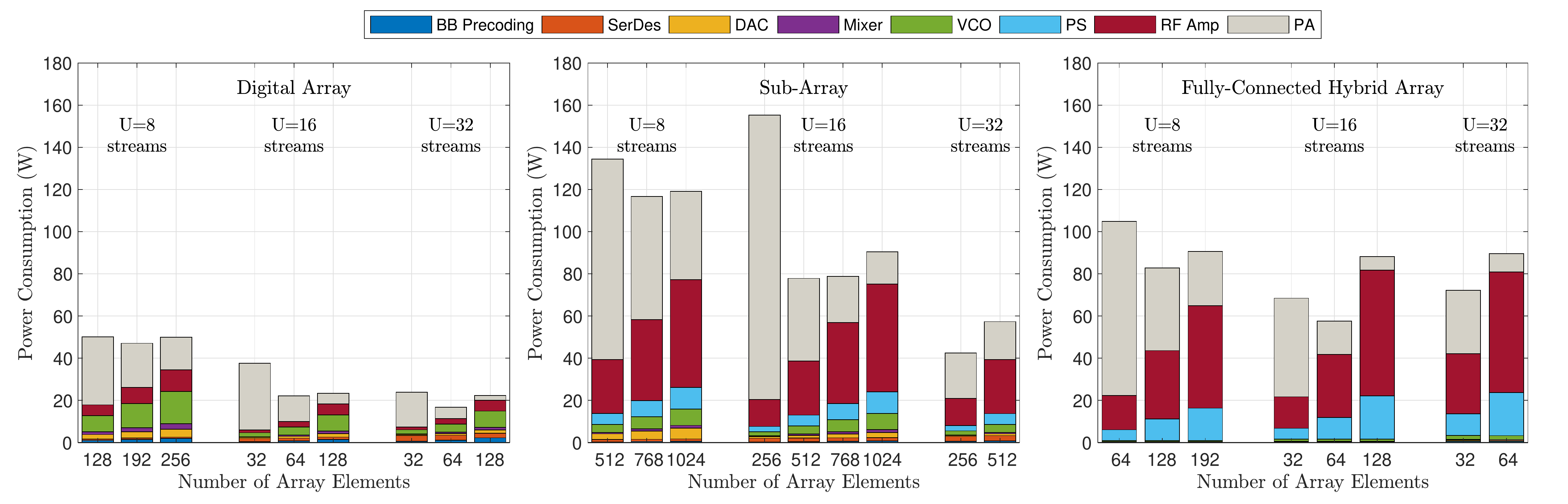}
\end{center}
\caption{Total power consumption for three architectures operating in the Dense Urban MBB use case. For each array architecture with varying array size, other design parameters are chosen according the analysis in Section~\ref{sec:design_parameters} and 58.8bps/Hz SE target demands are guaranteed in the corresponding LOS environment listed in Table~\ref{tab:link_budget_use_cases}.}
\label{fig:power_division_caseI}
\end{figure*}

\begin{figure*}[t]
\begin{center}
\includegraphics[width=1\textwidth]{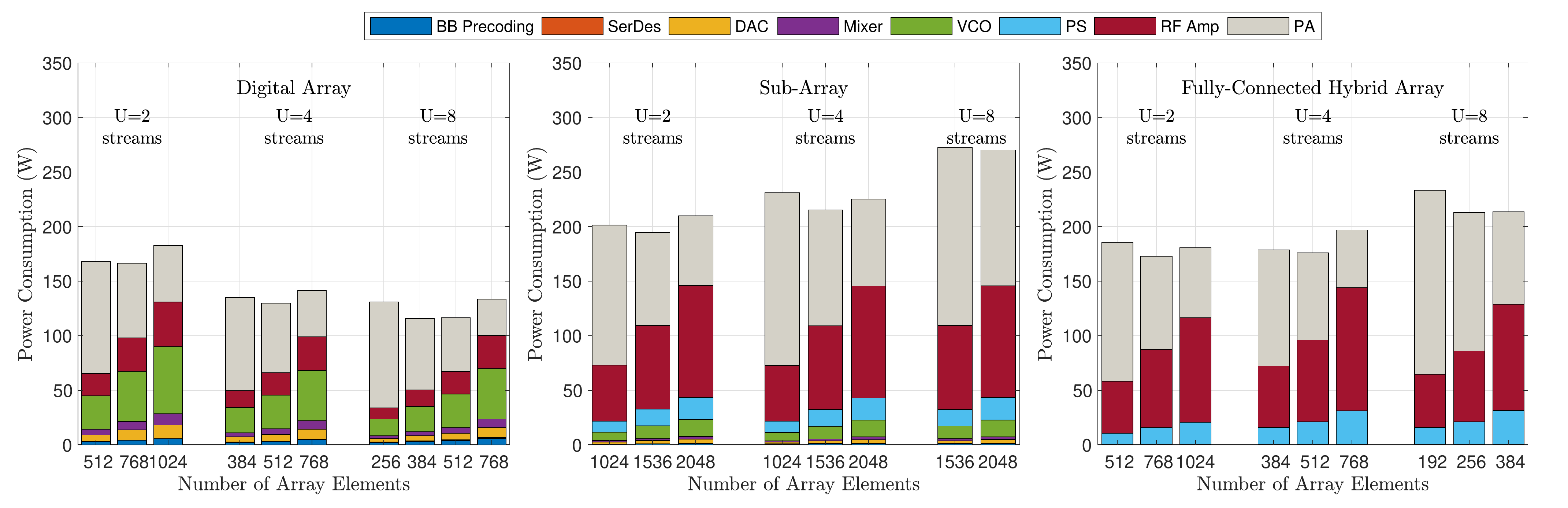}
\end{center}
\caption{Total power consumption for three architectures operating in the 50+Mbps Everywhere use case. For each array architecture with varying array size, other design parameters are chosen according the analysis in Section~\ref{sec:design_parameters} and 4.7bps/Hz SE target demands are guaranteed in the corresponding NLOS environment listed in Table~\ref{tab:link_budget_use_cases}.}
\label{fig:power_division_caseII}
\end{figure*}

In this section, we present the power and hardware cost comparison among three architectures. Then, we discuss the scalability of these architectures for future trends. Specifically, we focus on the impact of increased throughput requirement and improved energy efficiency in digital computation due to silicon scaling.

%
%
\subsection{Power consumption of mmW array architectures}

\begin{figure}
\begin{center}
\includegraphics[width=0.5\textwidth]{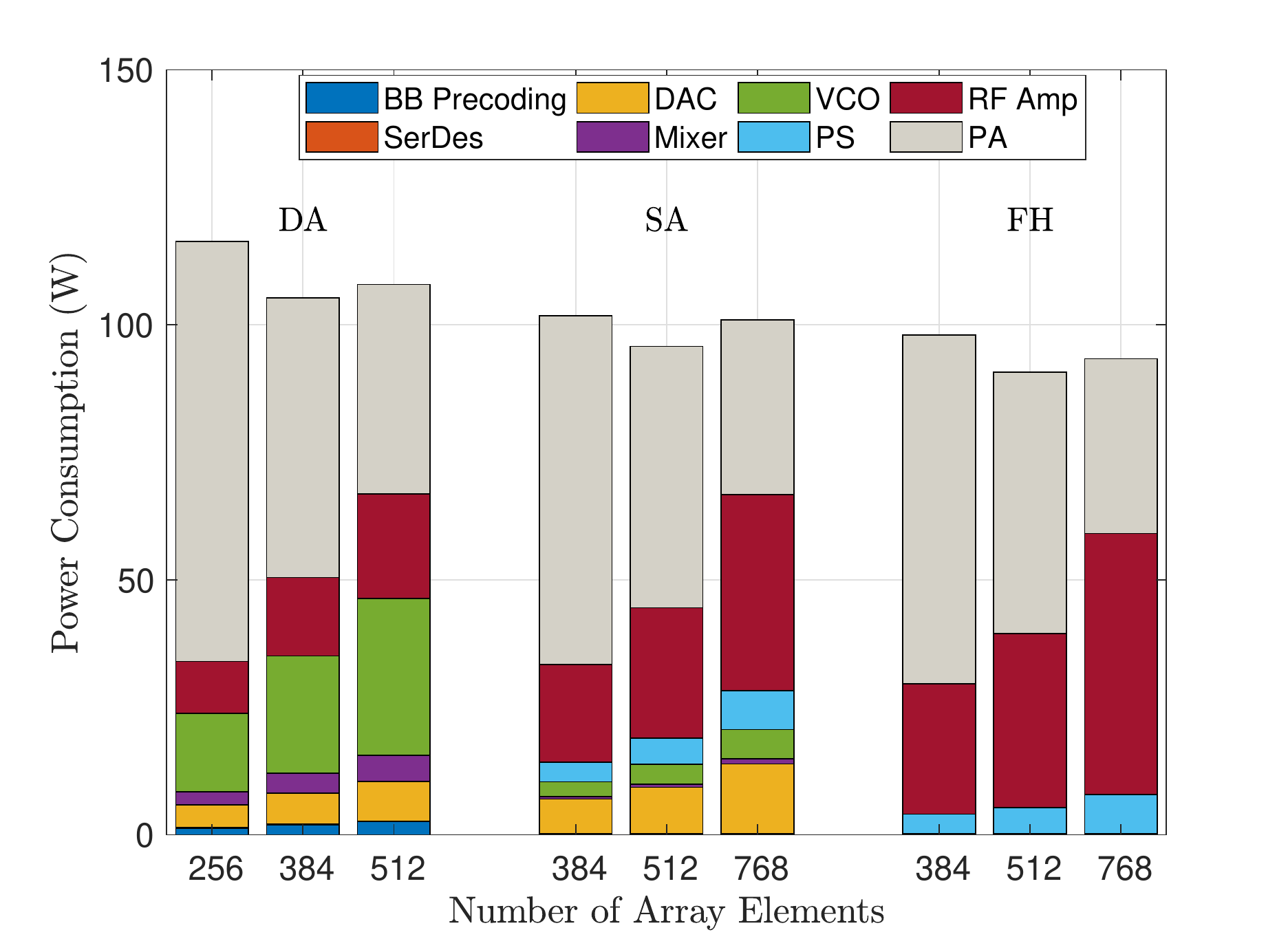}
\end{center}
\caption{Total power consumption for three architectures operating in the Self-backhauling use case. For each array architecture with varying array size, other design parameters are chosen according the analysis in Section~\ref{sec:design_parameters} and 11.8bps/Hz SE target demands are guaranteed in the corresponding LOS environment listed in Table~\ref{tab:link_budget_use_cases}.}
\label{fig:power_division_caseIII}
\end{figure}

The required power consumption in three use cases is presented in Figure~\ref{fig:power_division_caseI} to \ref{fig:power_division_caseIII}. All designs meet the SE requirement and the quantizations in DSP, SerDes, DAC, and PS are optimized. We observe that the system power consumption is a concave function of array size except few exceptions that will be discussed in later paragraphs. The concavity comes from the trade-off between PA power and processing power in other circuits blocks for different antenna array sizes. In the figures, the range of antenna element number $N$ for all scenarios is chosen to be close to green point, one that minimizes system power consumption.

Taking a closer look at Dense Urban MBB use case in Figure~\ref{fig:power_division_caseI}, we have the following conclusions.  
Firstly, DA and FH have similar green point of array size when the same number of streams $U$ is used, while green point of SA is much larger. This is due to the inefficiency of array gain (\ref{eq:maximum_gain}) when SA splits antenna with sub-groups. The exception occurs in SA with $U=32$ streams. When SA uses small antenna number and high multiplexing level, it effectively becomes a digital array. In fact, the green point for SA with $U=32$ streams occurs in $N=32$. It requires RF-chain to be connected with one antenna, and it makes SA a fully digital array. In the rest of comparison discussion, we focus on regime where each RF-chain is connected to $K_{\SA} = 8$ antennas and do not further consider regime for $N<256$ with $U=32$ streams.
Secondly, increasing $U$ reduces system power consumption in DA and SA. With the fixed $N$, increasing $U$ reduces required transmit power and thus saves DC power of PA. Besides, increasing $U$ does not require additional hardware resources except baseband precoding and SerDes throughput. With the benefits of quantization requirement reduction from Figure~\ref{fig:DAC_quantization} and high DSP efficiency, the negative impact of additional hardware resources is marginal. 
Thirdly, the transmit power and power consumption of PA reduces when FH uses higher $U$, but the system does not necessarily benefits. Part of the reason is that power in other circuits blocks linearly scales with stream number and they become system bottleneck in high-$U$ regime. Another important fact is that a power efficient design tends to reduce $N$ to save processing power when increased $U$. It implies FH needs to deal with higher interference from the increased beam-width. In fact, FH with $N=16$ cannot meet SE requirement when using $U=32$ beams. 
At last, comparing with the best designs of all architectures, we conclude that DA is the most power efficient architecture. The best design of SA becomes DA and the best design of FH still requires $240\%$ more power than DA. 

\begin{figure*}[t]
\begin{center}
\includegraphics[width=1\textwidth]{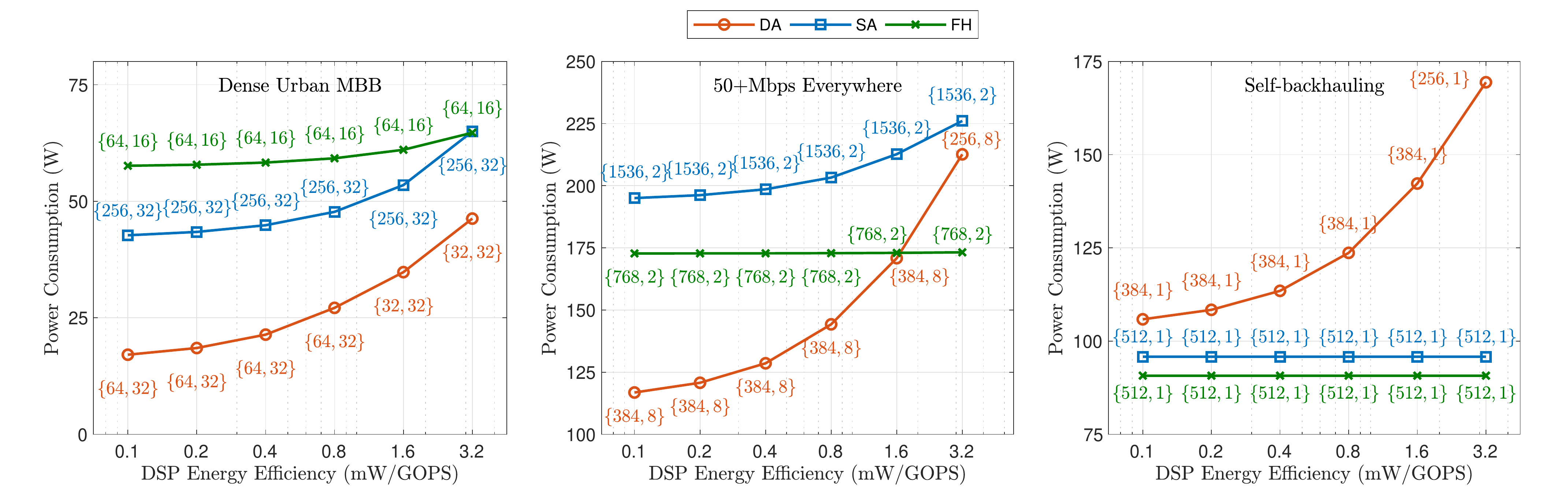}
\end{center}
\caption{System power consumption with different DSP energy efficiency in the unit of mW/GOPS. In all cases, optimal design parameters that reach SE target with lowest power consumption are chosen. The optimal antenna number $N$ of multiplexing level $U$ are labeled insider brackets $\{N,U\}$ which are adjacent to corresponding data markers of system power consumption.}
\label{fig:DSP_scaling}
\end{figure*}
The system power consumption in 50+Mbps Everywhere is shown in Figure~\ref{fig:power_division_caseII}. We have the following findings. 
Firstly, the benefits of using higher multiplexing are not as prominent as in MBB case. According to Section~\ref{fig:Tx_power_array_size_tradeoff} and corresponding analysis, it is mainly caused by smaller target SINR relaxation by reducing U. In fact, SA requires to use higher transmit power and thus DC power of PA. 
Secondly, large array size $N$ is required to for power efficient system. Overall, system requires more hardware and power consumption than in Dense Urban MBB and it implies the intrinsic disadvantage of mmW to provide ubiquitous connection even in small cell size. 
At last, DA remains the most efficient architecture and best design of hybrid architecture require nearly $50\%$ more power. This is a surprising result. One may expects that hybrid architectures outperforms DA when system is optimized for beamforming rather than multiplexing in this NLOS environment. With $U=2$, we do observe comparable power consumption. However, DA further reduces its power by levering on increasing $U$ with negligible additional processing power consumption. Hybrid architectures either requires higher transmit power, e.g., SA, or excessive processing power, e.g., FH, to increase $U$.

The only use case in our survey that hybrid architectures outperform DA is Self-backhauling where multiplexing level is limited due to point-to-point communication environment of LOS channel. In Figure~\ref{fig:power_division_caseII}, the DA requires $18\%$ more power as compared to hybrid architectures. This small power margin is due to the fact that the DA requires nearly 4 bits smaller quantization than hybrid architectures according to Figure~\ref{fig:DAC_quantization} and it prevents excessive power consumption in BB precoding, SerDes and DAC. Overall in this use case, the SA and FH have similar power consumption. In fact, SA and FH have the same the number of phase shifters when using same number of antenna elements. The difference between them lies in the power consumption of signal routing. The SA has more RF-chains than FH and therefore SA requires more power in high precision DAC and VCOs. The FH has only one RF-chain but it requires more power for RF signal distribution than SA.

In Figure~\ref{fig:power_division_caseI} to \ref{fig:power_division_caseIII}, DAC and BB precoding power has small proportion in the DA system, even when high multiplexing or large array size is used. Part of the reason is the ENOB requirement relaxation according to Section~\ref{sec:design_parameters}. A more important factor is the DSP energy efficiency. Our study is based the assumption that baseband processing is implemented on application-specific integrated circuits (ASIC). In deploying mmW DA, programmable DSP or Field-Programmable Gate Array (FPGA) based BB processor provides flexibility of reconfiguring BB precoding scheme, with the cost of order-of-magnitude more power consumption \cite{6982237}. In Figure~\ref{fig:DSP_scaling}, the system power of all architectures are compared when different DSP efficiencies are used. Throughout all cases, all design parameters are optimized such that lowest power consumed in reaches SE target, and the required array size $N$ and multiplexing level $U$ is labeled in the figure. We have the following findings. 
Firstly, DA is most sensitive to the decreased DSP efficiency. An efficient design would use smaller array size when BB precoding becomes bottleneck since it effectively reduces DSP burden. SA is less sensitive due to a much smaller number of RF-chains except in Dense Urban MBB where SA effective behaves as a digital array. FH is least sensitive to DSP efficiency.
Secondly, with 3.2mW/GOPS, a FOM that can be reached by reconfigurable digital processor using 90 to 130nm process \cite{6982237}, DA remains the best architecture in Dense Urban MBB. In the rest use cases, DA becomes less competitive in power consumption.


\subsection{IC Areas and cost of mmW array architectures}
\begin{figure}
\begin{center}
\includegraphics[width=0.5\textwidth]{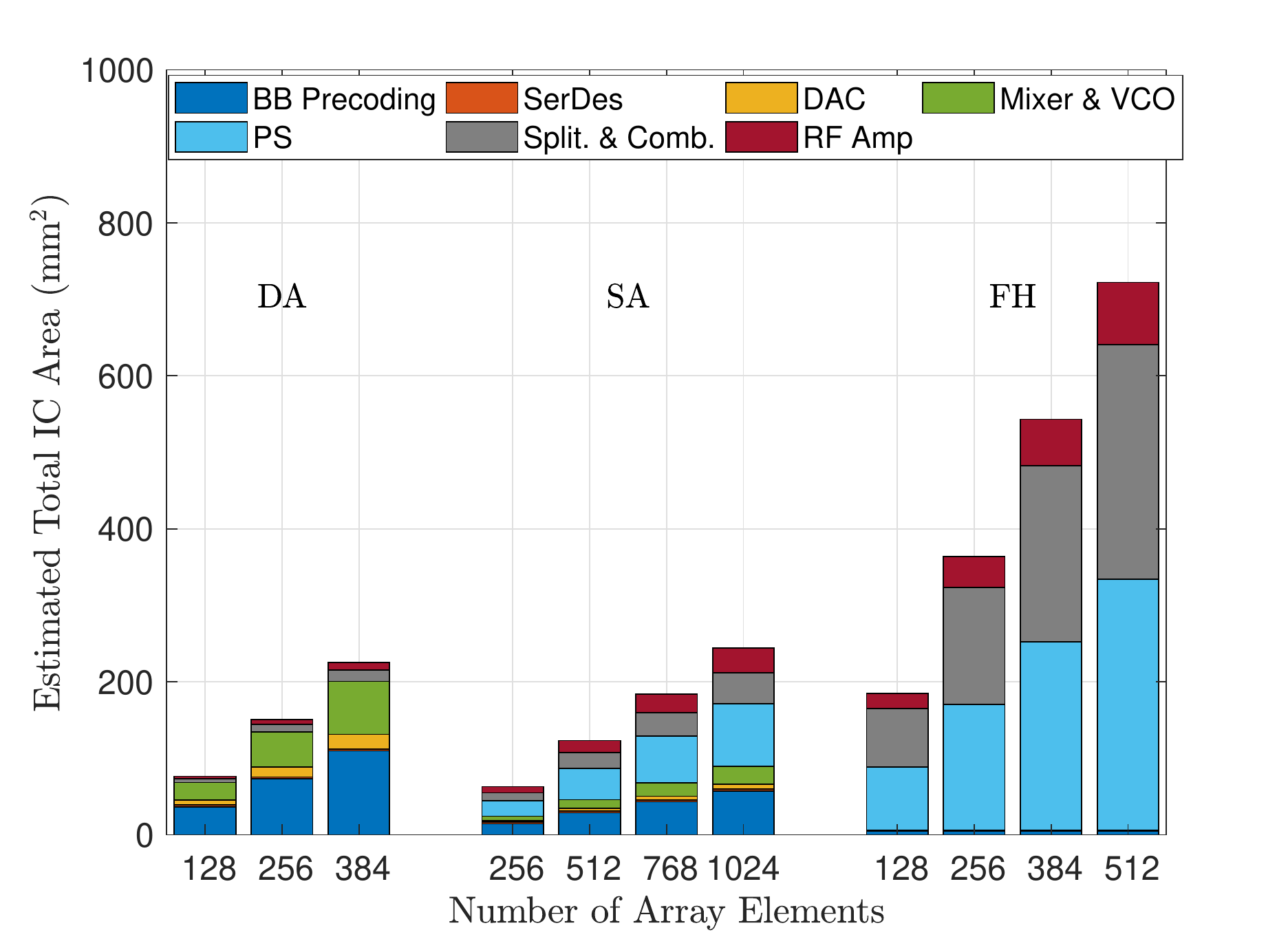}
\end{center}
\caption{IC area breakdown of three architectures.}
\label{fig:area_division}
\end{figure}

In Figure~\ref{fig:area_division}, the required IC area is presented as a function of array size. Note that increasing the multiplexing capability forces DA to have more powerful and larger DSP, and it also forces FH to have more RF-chain and complicated distribution network. Since maximum multiplexing of $U = 16$ does not significantly affect the optimal design for power consumption, we use $U=16$ for DA and FH while $U=32$ for SA. As shown in the figure, the largest contributor in DA is the DSP, which is expected to be further reduced so long as Moore Law reduces silicon area. SA remains competitive in IC area with DA. However, the cost of PA, which is likely to be fabricated with other material, is likely to require additional cost for SA due to the requirement of larger antenna number to be power efficient. FH requires the largest IC area due to the full connection nature between RF-chains and large number of antenna elements.

\section{Discussions on Open Research Challenges}

Admittedly, the power and IC area analysis for three array architectures provided are preliminary estimates based on the surveyed literature. In particular, the effect of the extra digital processing on power consumption and area depends on actual design and are hard to analyze at this point. Besides, some open research questions remain and were not covered in this paper. First one is the issues of synchronization among large number of array elements. In the centralized LO distribution architecture, each element re-generates clock from the same references but global LO distribution may not be area and energy efficient \cite{920580}. Under distributed LO scheme, independent LOs help reduce impact of phase noise \cite{7491244} but system needs to be calibrated periodically to avoid loss of coherency across elements. Second issue is related to compensation of PA nonlinearity. Digital predistortion (DPD) is important in massive transmitter array design. Conventionally, DPD is designed DA and DSP is implemented for each pair of transmitter chain and PA. Due to increased processing and power of DPD, the overall gains in power efficiency for large number antenna arrays need to be analyzed and optimized. DPD for SA \cite{8334305,8359454} and FH \cite{7952803} are actively investigated by researchers. Thirdly, other design variations, including phase-and-magnitude analog precoder and active RF splitter and combiner \cite{5556449} can help reduce the complexity and power consumption of the hybrid arrays. Last, our survey reveals the benefits of using high multiplexing level for power saving in the hardware. However, high multiplexing brings additional burden in higher layers of system, e.g., network layer faces more challenges to schedule users with non-overlapping propagation paths, and their impact needs to be incorporated in more comprehensive study.

In this work, we reveal that the conventional belief that hybrid array architecture is more cost and energy efficient than digital architecture is not necessarily true when comprehensive hardware block is modeled and system adopts optimized design parameters. Similar findings were reported for the receiver array during the period when this work is written \cite{Zorzi_Rx_array,another_rx_array}. It is worth noting that these works, including ours, focus on the additive uniformly distributed quantization error model and linear MIMO processing model. However, such quantization error model becomes less precise when data samples and quantization error are correlated, which occurs when data converters have significantly small number of bits. Besides, linear MIMO processing is not optimal. In fact, in the receiver array a variety of nonlinear combining and decoding algorithms are proposed, e.g., successive interference cancellation based combining \cite{7445130}, approximate message passing \cite{7355388}. Besides, the precision requirement of DAC and analog-to-digital (ADC) devices are strongly dependent on processing algorithms, e.g., algorithm tailored for 1-bit ADC \cite{7876856}. It remains open research question how to use advanced signal processing to further reduce power consumption and cost of mmW array. 

The Matlab code for simulation and data for system level power comparison is released in \cite{hyan_git} for readers that are interested in results with different design choices and hardware specifications.
%
%
\section{Conclusion}
\label{sec:Conclusion}
Building energy and cost efficient massive array is one of the major challenge in implementing and deploying mmW networks in the 5G era. In this work, we study and compare three array architecture candidates, digital architecture and two variation of analog-digital hybrid architectures and discuss various hardware design trade-offs. Specifically, the required power, IC area of circuits blocks are modeled as functions of key design parameters in each architecture. Based on the state-of-the-art circuits design and measurement results, we evaluate the power and IC area of circuits blocks. We compare three array architectures when the associated design parameters are optimized to meets the spectral efficiency targets in three representative 5G-NR use cases with the most efficient manner. The results show that digital architecture is the most efficient in power and area. The key intuition is that digital array can effectively save system power and area by levering on high multi-user multiplexing , which effectively reduces requirement of array size, transmit power, and hardware specifications in the RF-chains. The hybrid architectures require additional power to support more simultaneous spatial beams, either via additional transmit power to compensate for the loss array gain, or severely increased processing power. Besides, we reveal that the bottleneck of hybrid architectures are the RF signal distribution networks in their RF beamforming stages. 

%
%
\section{Acknowledgment}
This work is partially supported by National Science Foundation through grant 1718742. Authors would like to acknowledge Dr. Jefferey Lee and Dr. Zhao Yan for their helpful comments and discussions.

\appendix
\subsection{Required DAC quantization bits}
\label{appendix:DAC_ENOB}
In this subsection, we provide analysis of transmit noise $\sigma^2_{\text{n,tx}}$ in each architecture.

For each DAC, the quantization error is uniformly distributed in $[-A/2^B,A/2^B]$ where $A$ is the largest quantization level. Without signal cropping, $A$ depends on the peak-to-average-power-ratio (PAPR), i.e., $\text{PAPR} = A^2$ with unit signal power. The power of DAC quantization noise is
\begin{align}
\epsilon_{\text{DAC}}(B) 
=10\log_{10}\left[\frac{(2A)^2}{12(2^B)^2}\right] = 10\log_{10}(A^2/3) - 6B \text{ [dB]}.
\label{eq:DAC_noise}
\end{align}
Note that the above power is normalized with the input signal power of each DAC.

In DA architecture, the input signal power of DAC is amplified to $P_{\DA}^{(\text{out})}/N_{\DA}$. As a consequence, the transmitter noise power at output of each PA is $P_{\DA}^{(\text{out})}\epsilon_{\text{DAC}}(B_{\DA})/N_{\SA}$. With the uncorrelated\footnote{Correlation among quantization errors of DACs become non-negligible when quantization level is significantly small, e.g., one bit. Dithering is a technique to de-correlate them but is beyond the scope of this work.} quantization errors in each DAC, transmit noise is $\sigma^2_{\text{n,tx}} = P_{\DA}^{(\text{out})}\epsilon_{\text{DAC}}(B_{\DA})$.  

In SA architecture, due to the identical input signal of DACs in a virtual group, quantization noise remains the same as well. The quantization noises are coherent at the outputs of $N_{\SA}/U$ PAs within a virtual group and each has power $P_{\SA}^{(\text{out})}\epsilon_{\text{DAC}}(B_{\SA})/N_{\SA}$. As a consequence, the transmit noise is $\sigma^2_{\text{n,tx}} = P_{\SA}^{(\text{out})}N_{\SA}\epsilon_{\text{DAC}}(B_{\DA})/U^2$.

In FH architecture, the quantization noise from each DAC is amplified to $P_{\FH}^{(\text{out})}/(NU)$ in each PA. As a result, the total transmitter noise power is $\sigma^2_{\text{n,tx}} = P_{\FH}^{(\text{out})}N_{\FH}\sigma^2_{\text{DAC}}(B_{\FH})/U$.

\subsection{Impact of phase shifter quantization error and random error on beamforming gain}
\label{appendix:phase_shifter}
Consider a linear phased array system with $N$ antenna elements that steers a beam towards direction $\gamma$ in a 2D plane. Beamforming vector is given by $[e^{j\phi_1}, \cdots ,e^{j\phi_N}]$, where $\phi_n = (n-1)\pi\sin(\gamma)$. In the next, we derive beamforming gain at the main lobe for system with ideal and non-ideal phase shifters.

Let us denote the signal at the $n^{\text{th}}$ elements as $w_n$ with $|w_n| = 1/\sqrt{N},\forall n$ when all phase shifters are ideal. Clearly, the phase shifter needs to be set such that signals are constructively added in the intended direction, i.e., $w_n e^{\phi_n} = 1/\sqrt{N}$, and the beamforming gain is
\begin{align*}
G = \left|\sum_{n=1}^{N}w_ne^{j\phi_n}\right|^2 = N
\end{align*}

When all phase shifters are non-ideal, the signal at the $n^{\text{th}}$ element is denoted as $w^{\prime}_n = w_n \text{exp}(j\psi_n)$, where $\psi_n$ is the phase error due to quantization and random implementation impairment. With $Q$ bits quantization, the phase error $\psi_n$ is bounded as $|\psi_n| \leq \epsilon $ where $\epsilon = \pi/2^Q$. The corresponding beamforming gain is 
\begin{align*}
G^{\prime} = & \left|\sum_{n=1}^{N} (w^{\prime}_n e^{j \phi_n})\right|^2 = \left|\sum_{n=1}^{N} (w_n  e^{j\psi_n} ) e^{j \phi_n}\right|^2 \\
= & \frac{1}{N}\left|\sum_{n=1}^{N} e^{j\psi_n}\right|^2 =  \frac{1}{N}\left|\sum_{n=1}^{N}\cos(\psi_n)+ j\sum_{n=1}^{N}\sin(\psi_n)\right|^2\\
= &\frac{1}{N}\left[\sum_{n=1}^{N}\cos(\psi_n)\right]^2+\frac{1}{N} \left[\sum_{n=1}^{N}\sin(\psi_n)\right]^2 \\
\geq &\frac{1}{N}\left[\sum_{n=1}^{N}\cos\left(\psi_n\right)\right]^2 \\
\geq &N \cos^2\left(\epsilon\right)
\end{align*}
where the second inequality is valid so long as $Q\geq 1$, i.e., $|\psi_n|\leq\pi/2, \forall n$.

Therefore the gain reduction is bounded by
\begin{align*}
10 \log_{10} \left[\frac{G}{G^{\prime}}\right] \leq -20 \log_{10} \left[\cos\left(\frac{\pi}{2^Q}\right)\right] \text{[dB]}
\end{align*}
The above derivation shows that the gain drop in the main lobe is less than 0.68dB, 0.16dB and 0.04dB with $Q = 3 \text{ to }5$ bits quantization. Besides, these values are independent from the antenna size $N$. Equivalently, when phase shifter implementation error is less than $\epsilon = 22.5^{\circ}, 11.25^{\circ}$, and $5.625^{\circ}$, gain drop is also bounded by 0.68dB, 0.16dB and 0.04dB, respectively.
%
%



%
\bibliographystyle{IEEEtran}
\bibliography{IEEEabrv,references}

\end{document}